\documentclass[journal, twocolumn]{IEEEtran}

\usepackage{graphicx}
\graphicspath{{./figures/}}
\DeclareGraphicsExtensions{.pdf}
\usepackage{epstopdf}
\usepackage{subcaption}
\usepackage{multirow}
\usepackage{amsmath}
\usepackage[numbers]{natbib}
\usepackage{url}

\begin{document}

\title{Seismic Signal Denoising and Decomposition Using Deep Neural Networks}


\author{Weiqiang Zhu\thanks{zhuwq@stanford.edu}, S. Mostafa Mousavi and Gregory C. Beroza \\
  Department of Geophysics, Stanford University}

\maketitle

\begin{abstract}

Denoising and filtering are widely used in routine seismic-data-processing to improve the signal-to-noise ratio (SNR) of recorded signals and by doing so to improve subsequent analyses. In this paper we develop a new denoising/decomposition method, DeepDenoiser, based on a deep neural network. This network is able to learn simultaneously a sparse representation of data in the time-frequency domain and a non-linear function that maps this representation into masks that decompose input data into a signal of interest and noise (defined as any non-seismic signal). We show that DeepDenoiser achieves impressive denoising of seismic signals even when the signal and noise share a common frequency band. Our method properly handles a variety of colored noise and non-earthquake signals. DeepDenoiser can significantly improve the SNR with minimal changes in the waveform shape of interest, even in presence of high noise levels. We demonstrate the effect of our method on improving earthquake detection. There are clear applications of DeepDenoiser to seismic imaging, micro-seismic monitoring, and preprocessing of ambient noise data. We also note that potential applications of our approach are not limited to these applications or even to earthquake data, and that our approach can be adapted to diverse signals and applications in other settings. 

\end{abstract}

\begin{IEEEkeywords}
seismic denoising, decomposition, deep learning, convolutional neural network
\end{IEEEkeywords}

\section{Introduction}

Recorded seismic signals are inevitably contaminated by noise and non-seismic signals from various sources including: ocean waves, wind, traffic, instrumental noise, electrical noise, etc. Spectral filtering (usually based on the Fourier transform) is frequently used to suppress noise in routine seismic data processing; however, this approach is not effective when noise and seismic signal occupy the same frequency range. Moreover, selecting optimal parameters for filtering is non-intuitive, typically varies with time, and may strongly alter the waveform shape such that it degrades the analysis that follows.   

Due to these limitations, numerous efforts have been made to develop more effective noise suppression in seismic data e.g.~\citep{Abma-Claerbout1995, Oropeza2011, Bonar2012, Naghizadeh2012a, Liu2012f-x,Tian2014, zhu2015seismic, Chen-Fomel2015, Liu2015,Chen2016open, Chen2016sim, Huang2016damped, huang2016signal, Huang2017low, zhou2017spike, Zhou2017online, Chen2017empirical, huang2018regularized, Chen2018non}. Methods based on time-frequency denoising~\citep{donoho1994ideal,donoho1995adapting} form a large class of seismic denoising techniques. In this approach, noisy time series are first transformed into the time-frequency domain using a time-frequency transform, such as a wavelet transform ~\citep{Cao2005CWT, Gaci2014CWT, liu2016CWT, Mousavi2016Hybrid, Mousavi2016Custom, Mousavi2017CGV}, Short Time Fourier Transform (STFT)~\citep{Mousavi2016Adaptive}, S-transform~\citep{tselentis2012S-transform}, curvelet transform~\citep{Hennenfent2006Curvelet, Neelamani2008Curvlet, Tang2011Curvelet}, dreamlet transform~\citep{Wang2015Dreamlet}, contourlet transform~\citep{Shan2009contourlet}, shearlet transform~\citep{Zhang2018Shearlet}, empirical mode decomposition~\citep{Liu2014EMD, Chen2014EMD, Bekara-Baan2009EMD, Chen2014EMD, Chen2017EMD, Han-Baan2015EMD}, etc. The resulting time-frequency coefficients are modified (thresholded) to attenuate the coefficients associated with noise and to find an estimate of the signal coefficients. The modified coefficients are inverse transformed back into the time domain to reconstruct the denoised signal. 
The basic idea is to promote sparsity by transforming seismic data to other domains where the signal can be represented by a sparse set of features so that signal and noise can be separated more easily. 

These methods can suppress the noise even when it occupies the same frequency range as the signal, however, the choice of a suitable thresholding function to map the noisy data into optimally denoised signal can be challenging. Denoising performance of time-frequency methods can be improved in two ways: either by using a more effective sparse representation of the data, or by using a more flexible and powerful mapping function. Machine learning techniques, such as dictionary learning, have been used to improve seismic denoising by learning better sparse representations~\citep{Chen2016Dic, Chen2017Dic, siahsar2017Dic, Liu2018Dic}. The focus of this paper is on improving both sparsity and the mapping function using deep learning. Deep learning~\citep{LeCun2015c, Goodfellow2016b} is a powerful machine learning technique that can learn extremely complex functions through neural networks. Deep learning has been shown to be a powerful tool for learning the characteristics of seismic data~\citep{devries2018deep, perol2018convolutional,mousavi2018cred,  ross2018p, ross2018generalized, ross2018phaselink, zheng2017automatic, zhu2018phasenet}.

In this paper we present DeepDenoiser, a novel time-frequency denoising method using deep neural networks. This network is able to simultaneously learn a sparse representation of the input data and a high-dimensional non-linear function that maps this representation into desired masks from the training data set. Given an input data, DeepDenoiser produces two individual masks, one for seismic signal and the other for noise signal. The masks are further used to extract the corresponding waveforms from the input data. We use earthquake seismograms manipulated with various types of noise and non-earthquake signals recorded by seismic stations to train the network and demonstrate its performance. We apply the method to unseen noisy seismograms to illustrate its generalizability, to compare its performance with other denoising methods, and to document its ability to improve earthquake detection results. 

 \section{Method}

In the time-frequency domain, we represent recorded data, $Y(t, f)$, as the superposition of seismic signal, $S(t, f)$, and some additive natural/instrumental noise or non-seismic signals, collectively termed noise, $N(t, f)$:
\begin{equation}
Y(t,f) = S(t,f)+N(t,f)
\end{equation}
The objective of denoising is to estimate the underlying seismic signal (i.e., the denoised signal), $\hat{S}(t, f)$, from its noise contaminated version that minimizes the expected error between the true and estimated signal:
\begin{equation}
error = E||\hat{S}(t, f)-S(t, f)||_2^2
\end{equation}
where $\hat{S}(t, f) = {TFT}^{-1} \big\{ M(t,f) Y(t,f) \big\}$, ${TFT}^{-1}$ is the inverse time-frequency transform, $Y(t,f)$ is the time-frequency representation of noisy data, and $M(t,f)$ is a function that maps $Y(t,f)$ to a time-frequency representation of the estimated signal. 
\citet{donoho1994ideal} showed that this mapping can be carried out through a simple thresholding in a sparse representation where the thresholding value can be estimated from noise level assuming a Gaussian distribution.  

Here, we cast the problem as a supervised learning problem where a deep neural network will learn a sparse representation of the data to generate an optimal mapping function based on training samples of signal and noise data distribution. This is analogous to Wiener deconvolution as follows. We define our mapping functions as two individual masks, $M_S(t, f)$ and $M_N(t, f)$ for signal and noise respectively:
\begin{equation}
M_S(t,f) =\left[\frac{1}{1 + \frac{|N(t,f)|}{|S(t,f)|}}\right] 
\end{equation}
\begin{equation}
M_N(t,f) =\left[\frac{\frac{|N(t,f)|}{|S(t,f)|}}{1 + \frac{|N(t,f)|}{|S(t,f)|}}\right] 
\end{equation}
Each mask has the same size as the input time-frequency representation, $Y(t, f)$, and contains values between 0 and 1 that attenuate either noise or signal in  time-frequency space.   

\begin{figure*}
  \centering
  \includegraphics[width=\textwidth]{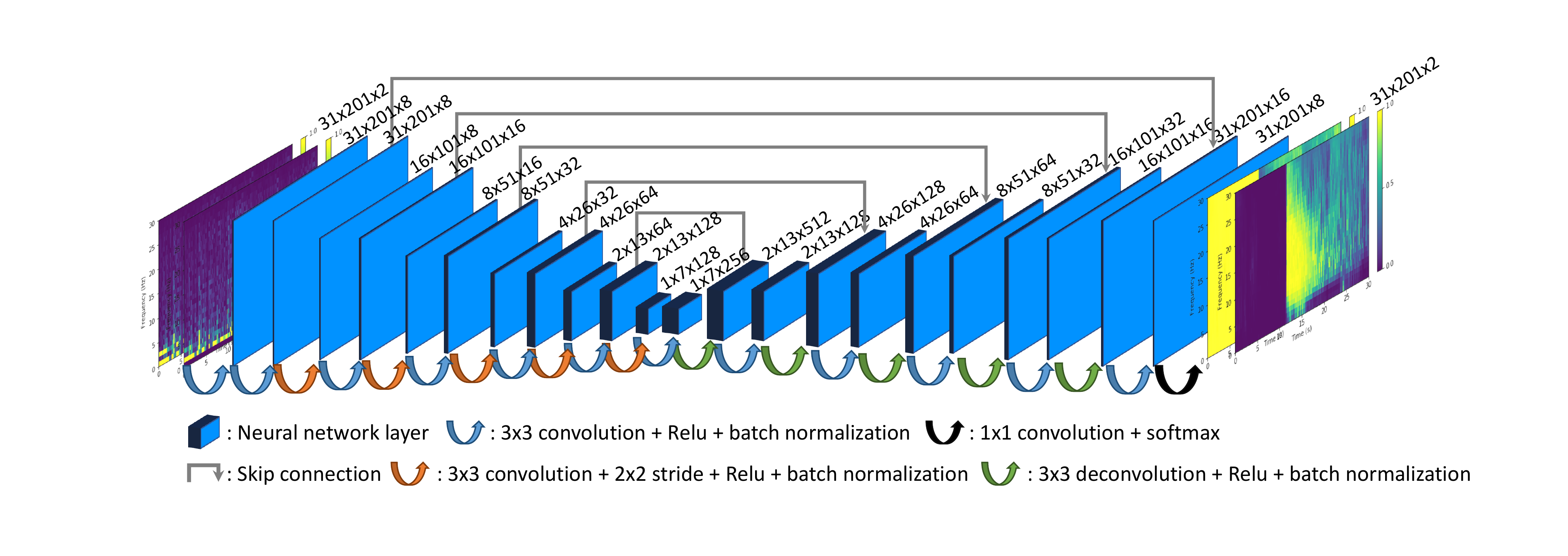}
  \caption{Neural network architecture. Inputs are the real and imaginary parts of the time-frequency representation of noisy data. Outputs are two masks for signal and noise extraction. Blue rectangles represent layers inside the neural network. The dimension of each layer is presented above it, and contains "frequency bins $\times$ time points $\times$ channels". Arrows represent different operations applied to layers. The input data first go through 3$\times$3 convolution layers with 2$\times$2 strides for down-sampling and then go through deconvolution layers~\citep{Noh2015} for up-sampling. Batch normalization and skip connections are used to improves convergence during training. The softmax normalized exponential function is applied in the last layer to predict the masks for signal and noise.}
  \label{fig:network}
\end{figure*}

Inspired by the capability of auto-encoders in learning a sparse representation of data with respect to an optimization objective, we designed our network in the form of a series of fully convolutional layers with descending and then ascending sizes (Fig.~\ref{fig:network}). Following \citet{Ronneberger2015}, we use skip connections to improve the convergence of training and prediction performance. 

The inputs to the first layer are the imaginary and real parts of the time-frequency coefficients of the data, $Y(t, f)$. In the last layer masks of signal and noise ($M_S(t, f)$ and $M_N(t, f)$) are provided as labels for training. The input time-frequency coefficients are processed and transformed through a series of 2D convolutional layers with a ReLU (rectified linear unit) activation layer and batch normalization~\citep{Ioffe2015}. The convolution filter size is kept constant (3$\times$3), but the feature space in the first half of the network is gradually shrunk using strides of 2$\times$2. These layers act as an effective feature extractor that can accelerate learning of a very sparse representation of input data at the bottleneck layer. In the second half of the network, deconvolution (transpose convolution layers) are used to generate a high-dimensional non-linear mapping of this sparse representation into output masks. In the last layer, a softmax normalized exponential function is used to produce masks. Through the training process the network learns both how to construct a sparse representation of data and optimal masks to separate signal from noise by optimizing a loss function (cross-entropy loss function).  

\begin{figure*}
  \centering
  \includegraphics[width=\textwidth]{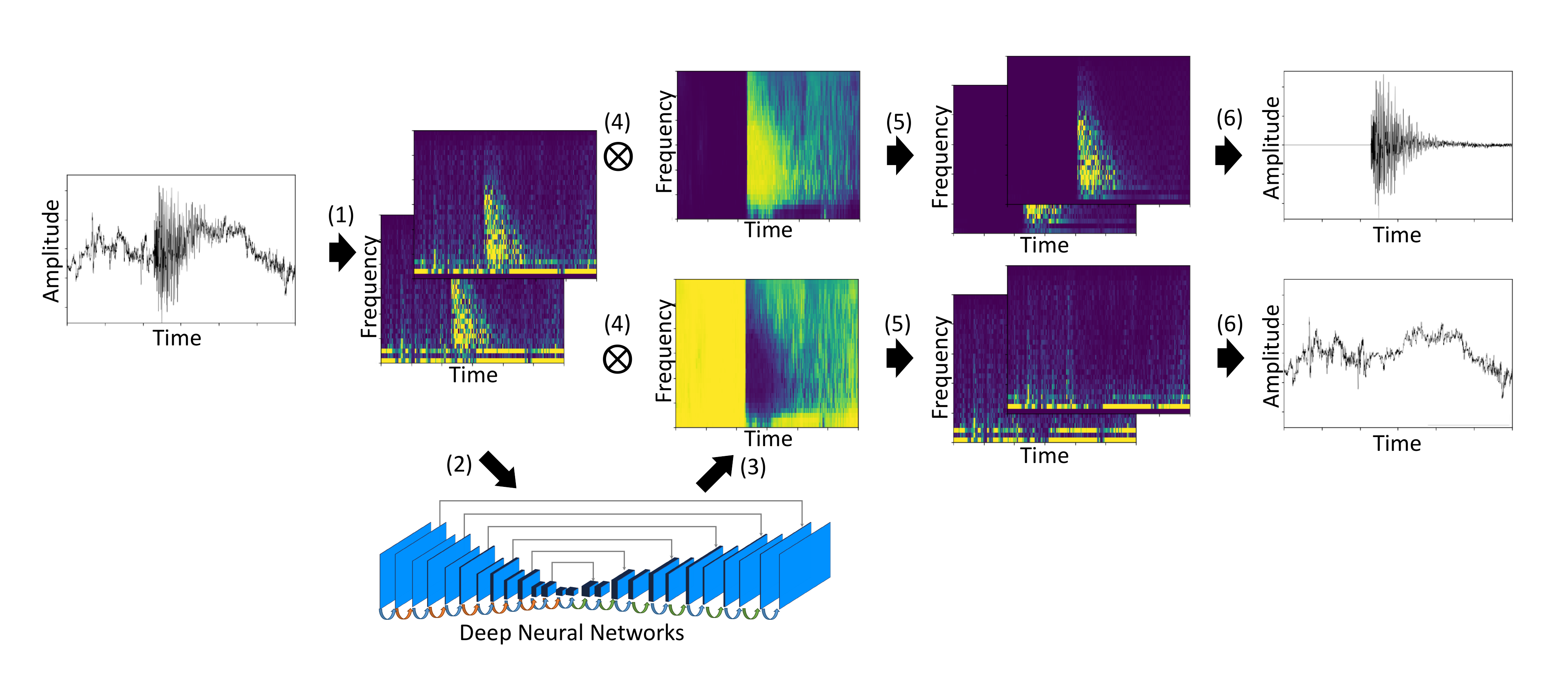}
  \caption{Data flow diagram of our denoising method. (1) The noisy data is transformed into the time-frequency domain using Short Time Fourier Transform (STFT). (2) The real and imaginary parts of time-frequency coefficients are fed into our deep neural network. (3) The neural network produces two masks for signal and noise based on input data. (4, 5) The associated masks are applied to the noisy-signal coefficients to estimate the time-frequency coefficients of the seismic signal and noise. (6) The denoised signal and noise in time domain are obtained using inverse STFT.}
  \label{fig:method}
\end{figure*}

Fig.~\ref{fig:method} shows the data-flow diagram of our method. First, the seismic waveform is transferred into the time-frequency domain. The trained network takes the real and imaginary parts of time-frequency coefficients as the input and produces individual masks for both signal and noise as the outputs. The masks are the targets in optimizing the neural network during training. The estimated time-frequency coefficients of the seismic signal, $\hat{S}(t,f)$, and noise, $\hat{N}(t,f)$, are obtained by applying the associated mask to the imaginary and real parts of the data coefficients, $Y(t, f)$. The final denoised signal and noise are obtained after inverse transforming $\hat{S}(t,f)$ and $\hat{N}(t,f)$ back into the time domain. Instead of defining different features and thresholds manually to enhance signal and attenuate noise, DeepDenoiser automatically learns richer features from semi-real seismic data that allows it to separate signal and noise in the time-frequency domain. 
Deep learning has the potential to provide a more effective and accurate automatic denoising tool for seismic data preprocessing, which can be applied to challenging tasks such as micro-earthquake detection.

\section{Network Training}
We use 30-second seismograms recorded by the high broadband channels (HN*) of the North California Seismic Network to train the neural network and test its performance. The dataset consists of 56,345 earthquake waveforms with very high signal-to-noise ratios (SNRs) as the signal samples and 179,233 seismograms associated with various types of non-earthquake waveforms as the noise samples. Both the signal and noise datasets are randomly split into training, validation and test sets. To form 'noisy' seismograms for training the neural network, we iterate through the signal training set repeatedly and in each iteration we add a randomly selected noise sample from the noise training set to the selected seismic signal to generate data of different SNR levels. The Short Time Fourier Transform (STFT) is applied to produce the time-frequency representation of the noisy waveforms. The 2D time-frequency matrices of the noisy waveforms are the input to our deep neural network. The real and imaginary parts are fed to the neural network as two separate channels so that the network is able to learn from both the time and phase information. The prediction targets of the neural network are two masks: one each for signal and noise, composed with equation (3) and (4). The same procedure is used to generate validation and test sets. The validation set is only used for fine-tuning the hyper-parameters of the network. This helps to identify and prevent over-fitting. The test set is used for the final results in this paper.

\section{Results}

\subsection{Test Set}
\begin{figure*}
\centering
\begin{subfigure}{0.41\textwidth}
\centering
  \includegraphics[width=\linewidth]{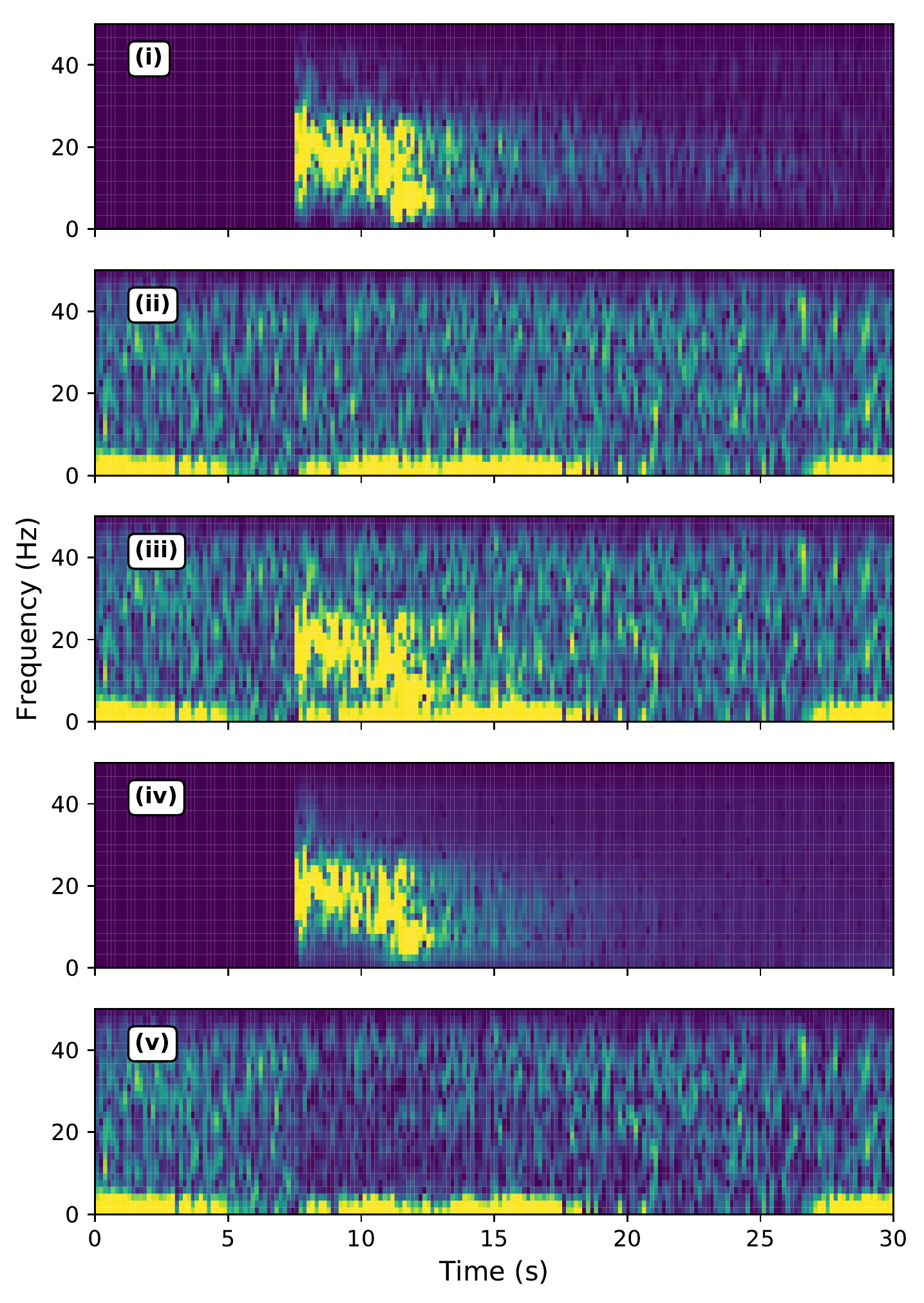}
  \caption{}
\end{subfigure}
\begin{subfigure}{0.41\textwidth}
\centering
  \includegraphics[width=\linewidth]{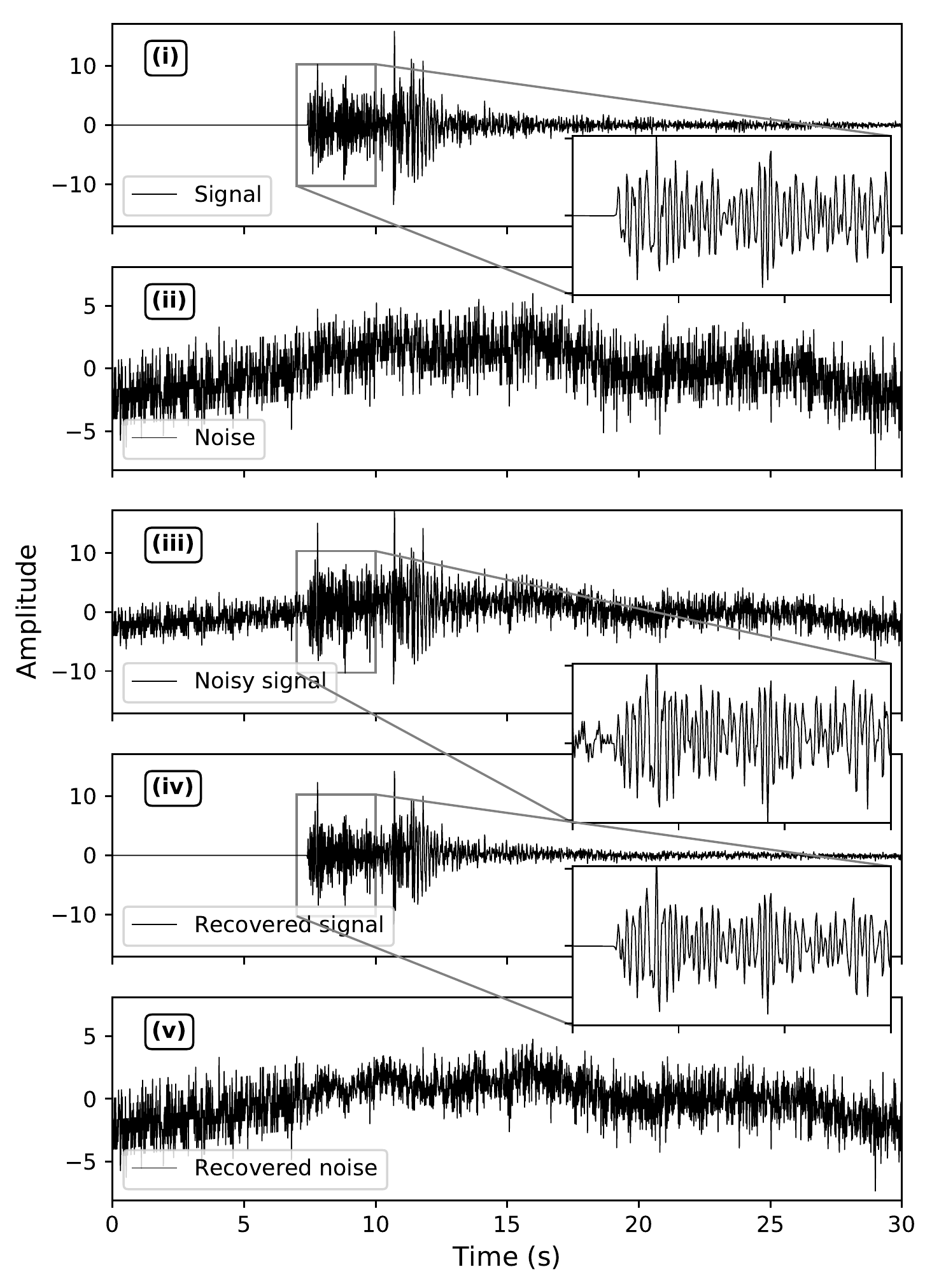}
  \caption{}
\end{subfigure}
\begin{subfigure}{0.41\textwidth}
\centering
  \includegraphics[width=\linewidth]{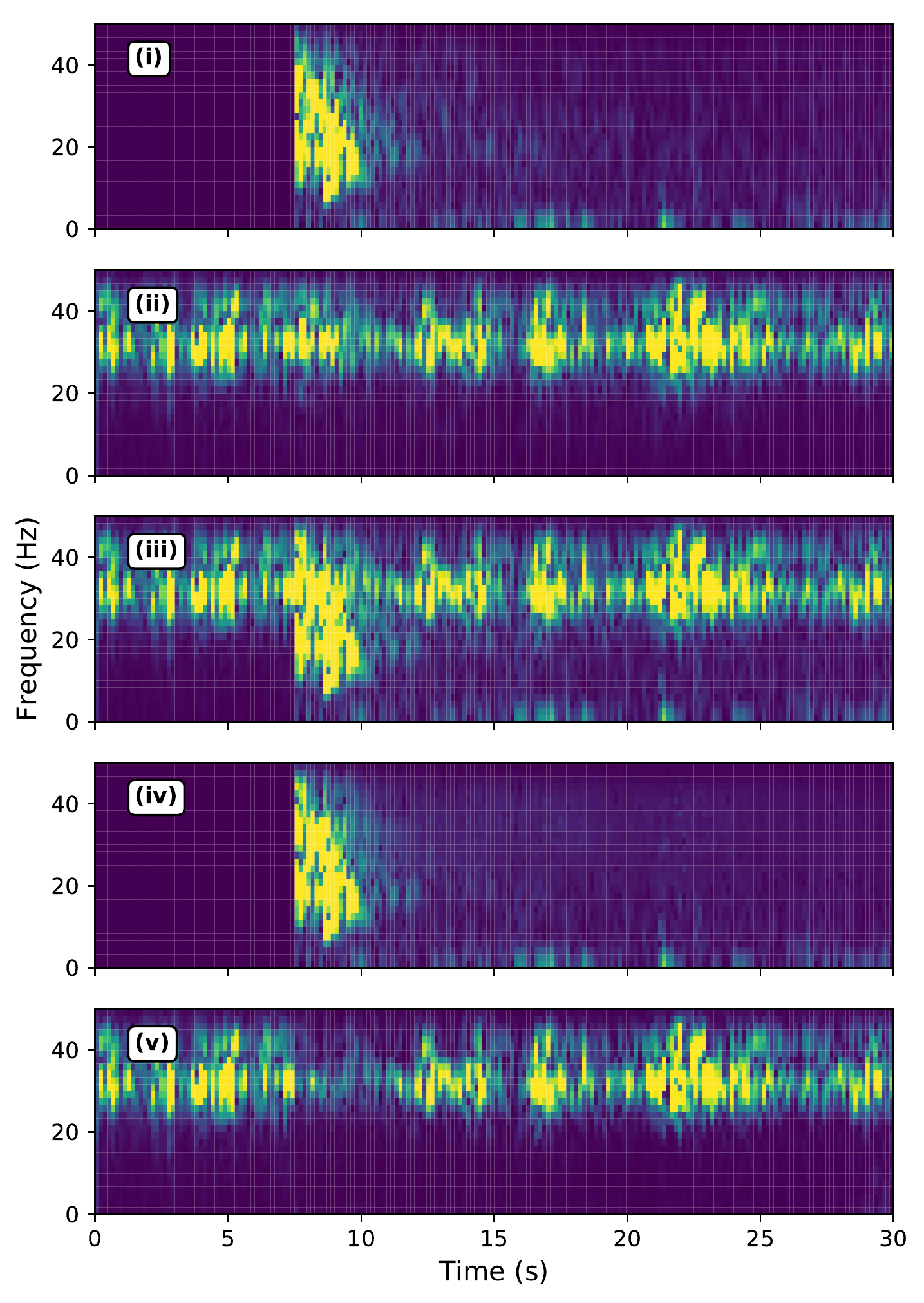}
  \caption{}
\end{subfigure}
\begin{subfigure}{0.41\textwidth}
\centering
  \includegraphics[width=\linewidth]{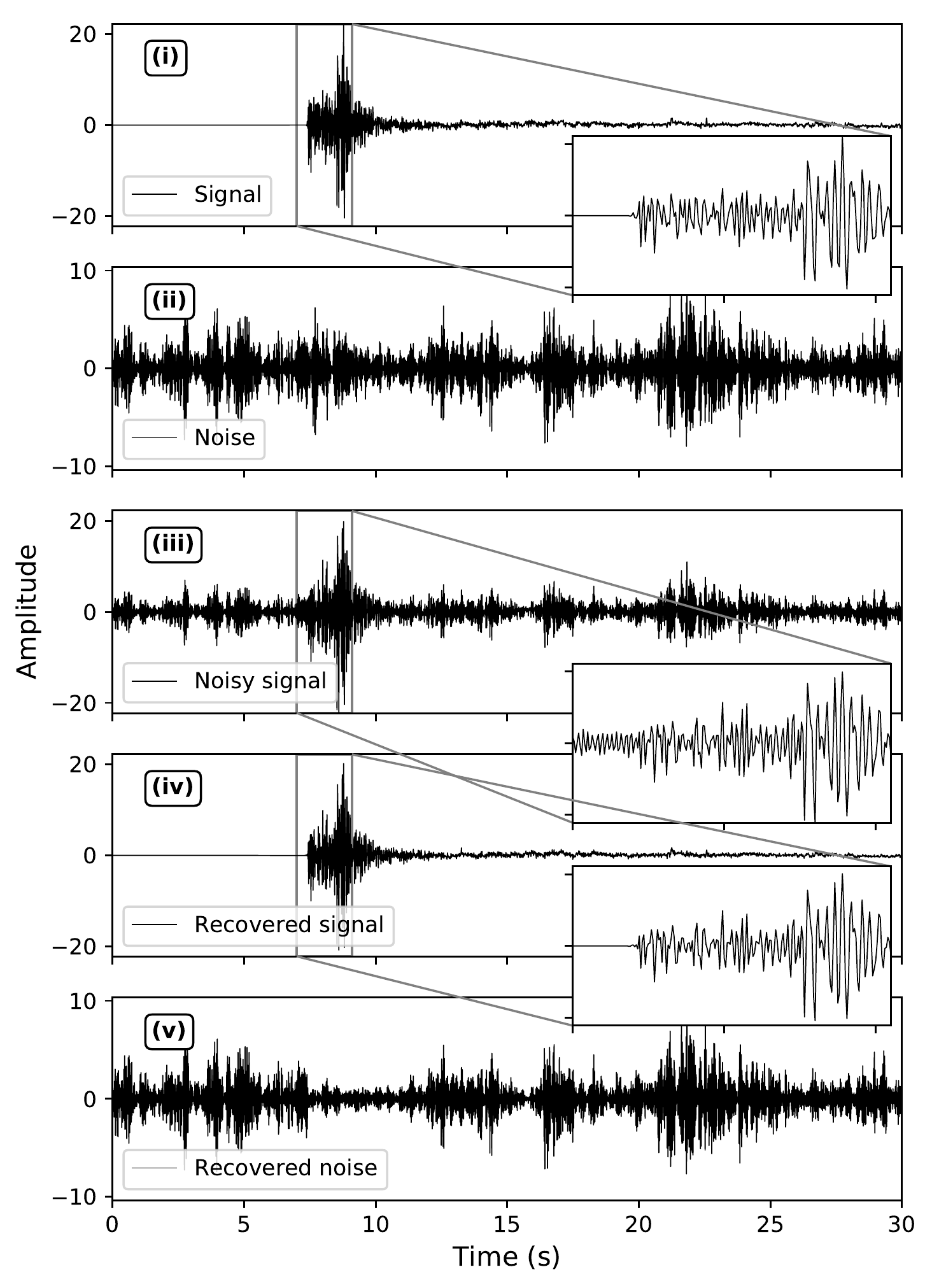}
  \caption{}
\end{subfigure}
\caption{Denoising examples: (a, c) time-frequency domain; (b, d) time domain. The time-frequency coefficients and waveforms of "clean" signal, real noise, and noisy signal are plotted in panels (i) (ii) (iii). Panels (a, c)(iv) show the denoised signal in the time-frequency domain, and panels (b, d)(iv) show its time domain waveform. The recovered noise is shown in panels (a, c)(v) and (b, d)(v).}
\label{fig:signal01}
\end{figure*}

We use the test set to analyze the final performance and visualize the denoising results. Fig.~\ref{fig:signal01} demonstrates that the network can successfully decompose the noisy inputs with different characteristics into denoised signal and noise. 
The algorithm can recover denoised signal with high accuracy (Fig.~\ref{fig:signal01}(b, d)(iv)). 
As can be seen from the same example, signal leakage is minimal and the waveform shape, frequency content, and amplitude characteristics are well preserved after denoising. These characteristics hold for the extracted noise as well (Fig.~\ref{fig:signal01}(b, d)(v)).  

\begin{figure*}
\centering
\begin{subfigure}{0.41\textwidth}
\centering
  \includegraphics[width=\linewidth]{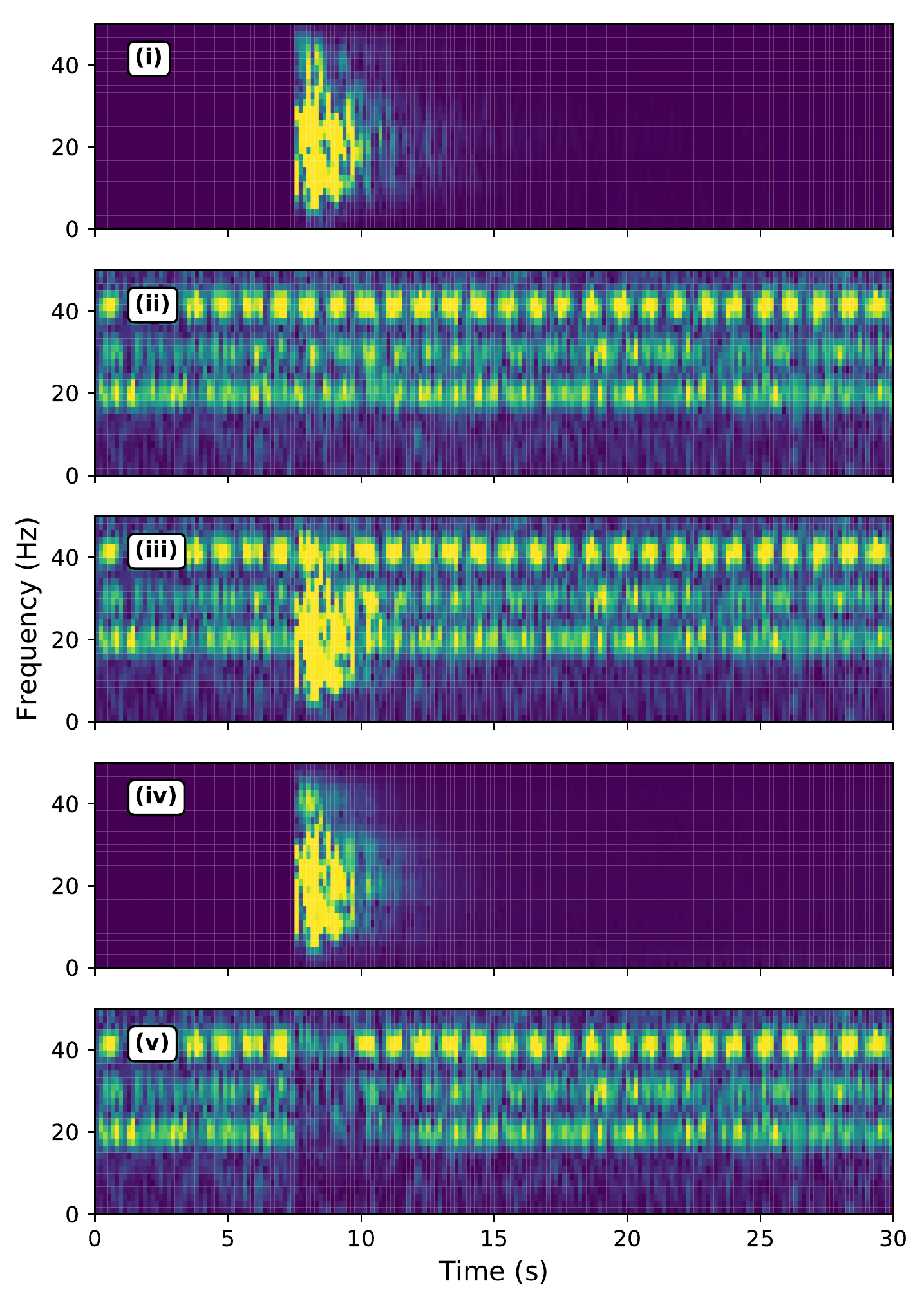}
  \caption{}
\end{subfigure}
\begin{subfigure}{0.41\textwidth}
\centering
  \includegraphics[width=\linewidth]{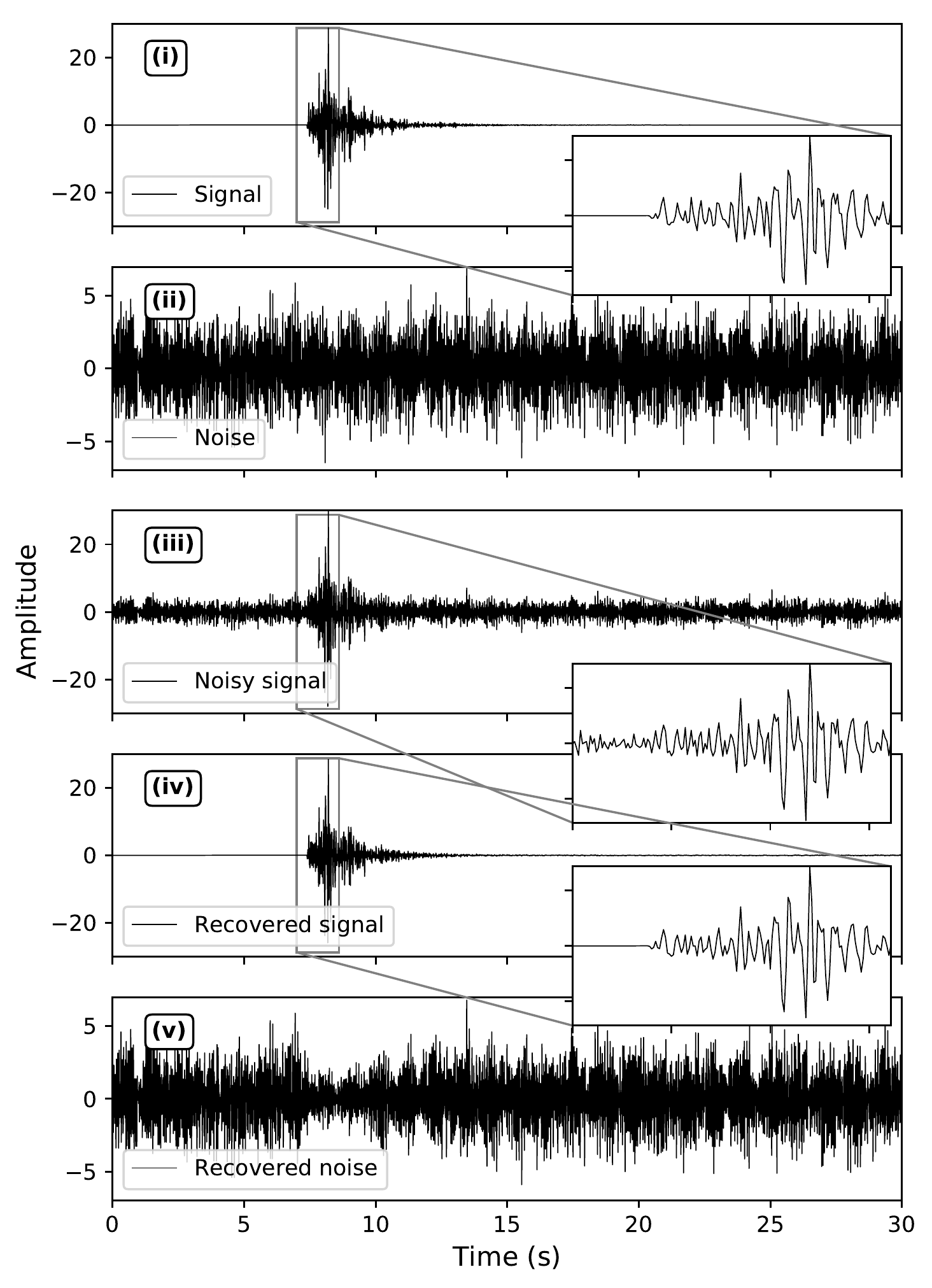}
  \caption{}
\end{subfigure}
\begin{subfigure}{0.41\textwidth}
\centering
  \includegraphics[width=\linewidth]{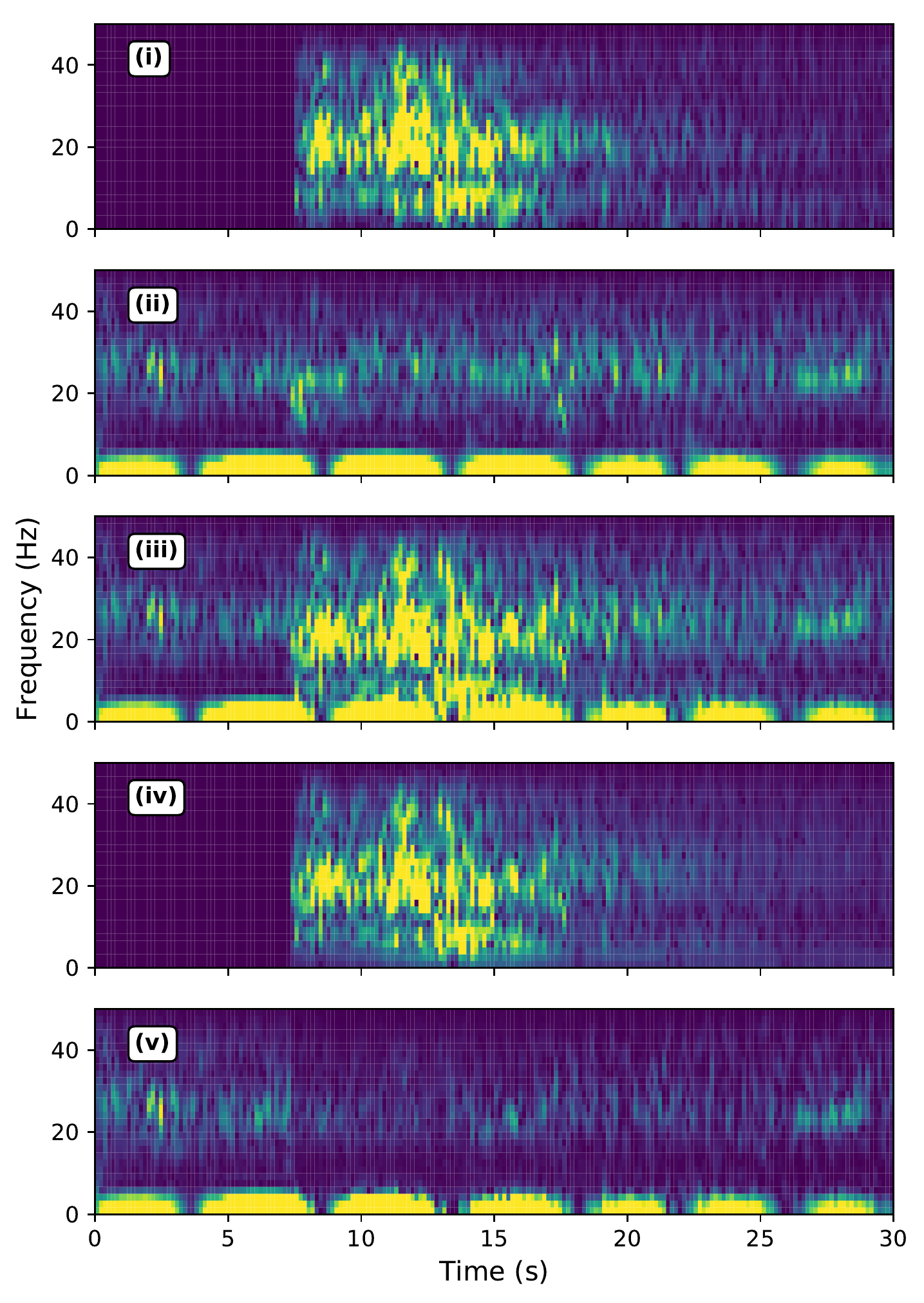}
  \caption{}
\end{subfigure}
\begin{subfigure}{0.41\textwidth}
\centering
  \includegraphics[width=\linewidth]{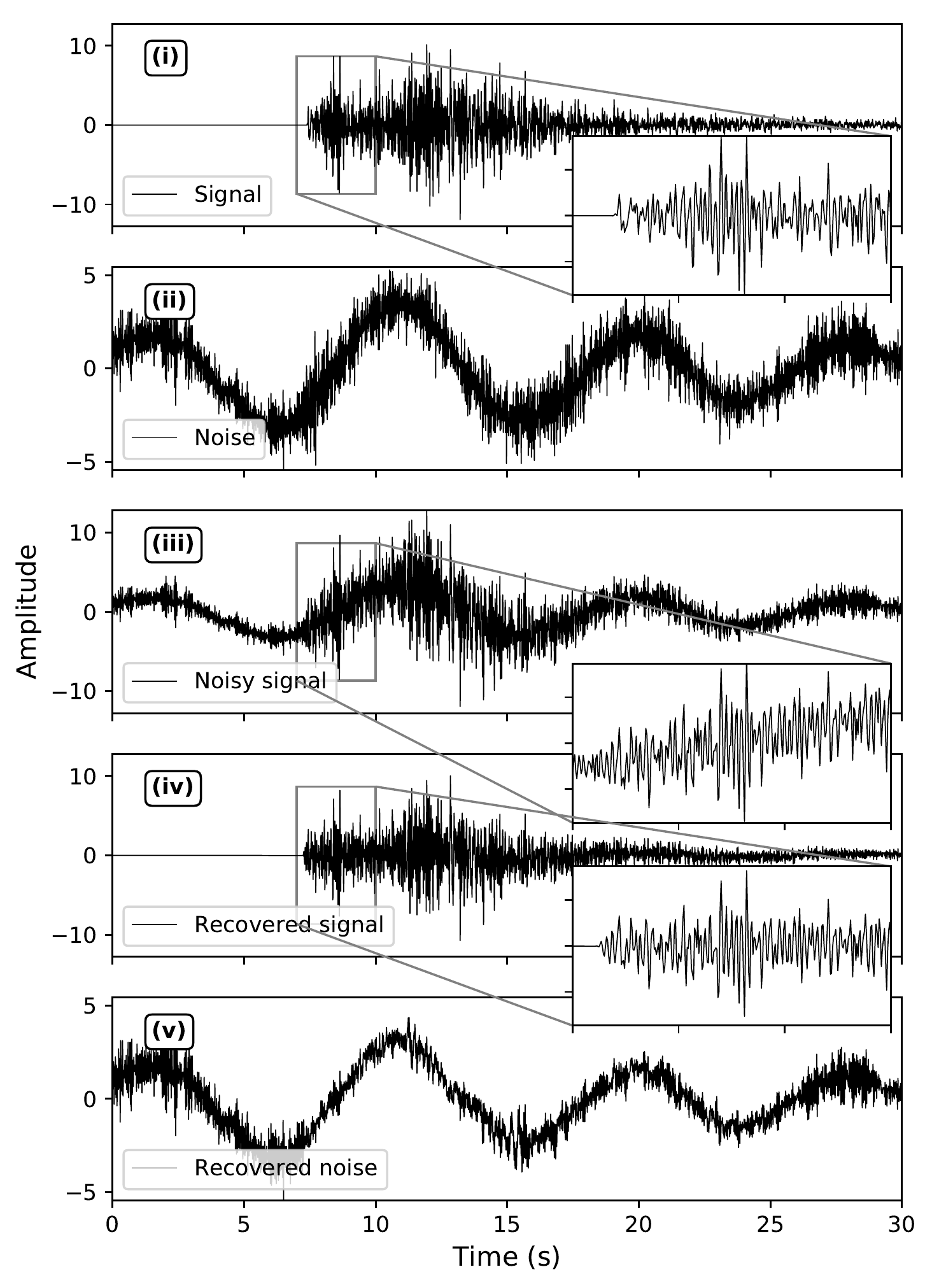}
  \caption{}
\end{subfigure}
\caption{Denoising performance in presence of strong colored noise. It is plotted in a same manner as Fig.~\ref{fig:signal01}.}
\label{fig:band_noise01}
\end{figure*}

\begin{figure*}
\centering
\begin{subfigure}{0.41\textwidth}
\centering
  \includegraphics[width=\linewidth]{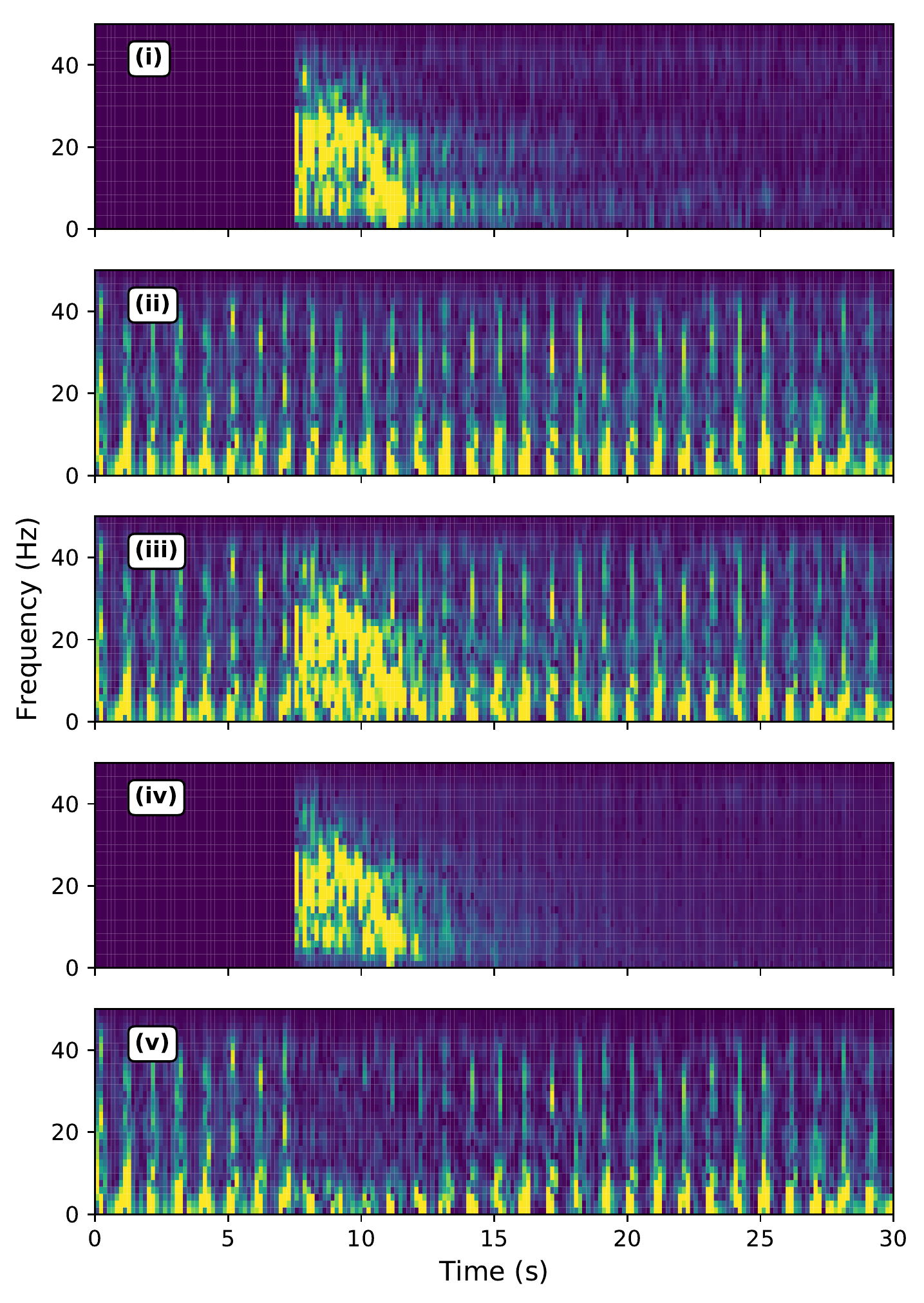}
  \caption{}
\end{subfigure}
\begin{subfigure}{0.41\textwidth}
\centering
  \includegraphics[width=\linewidth]{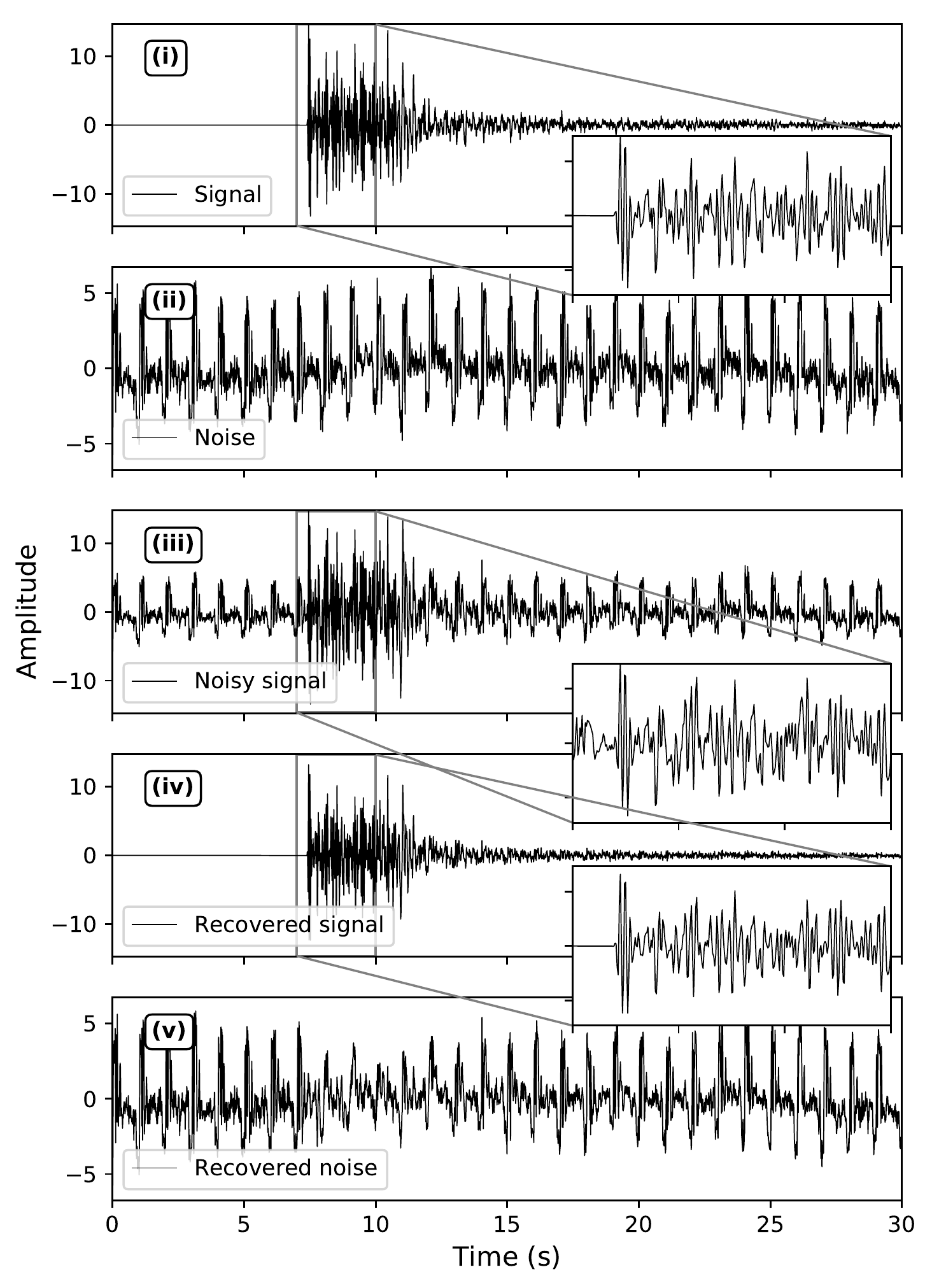}
  \caption{}
\end{subfigure}
\begin{subfigure}{0.41\textwidth}
\centering
  \includegraphics[width=\linewidth]{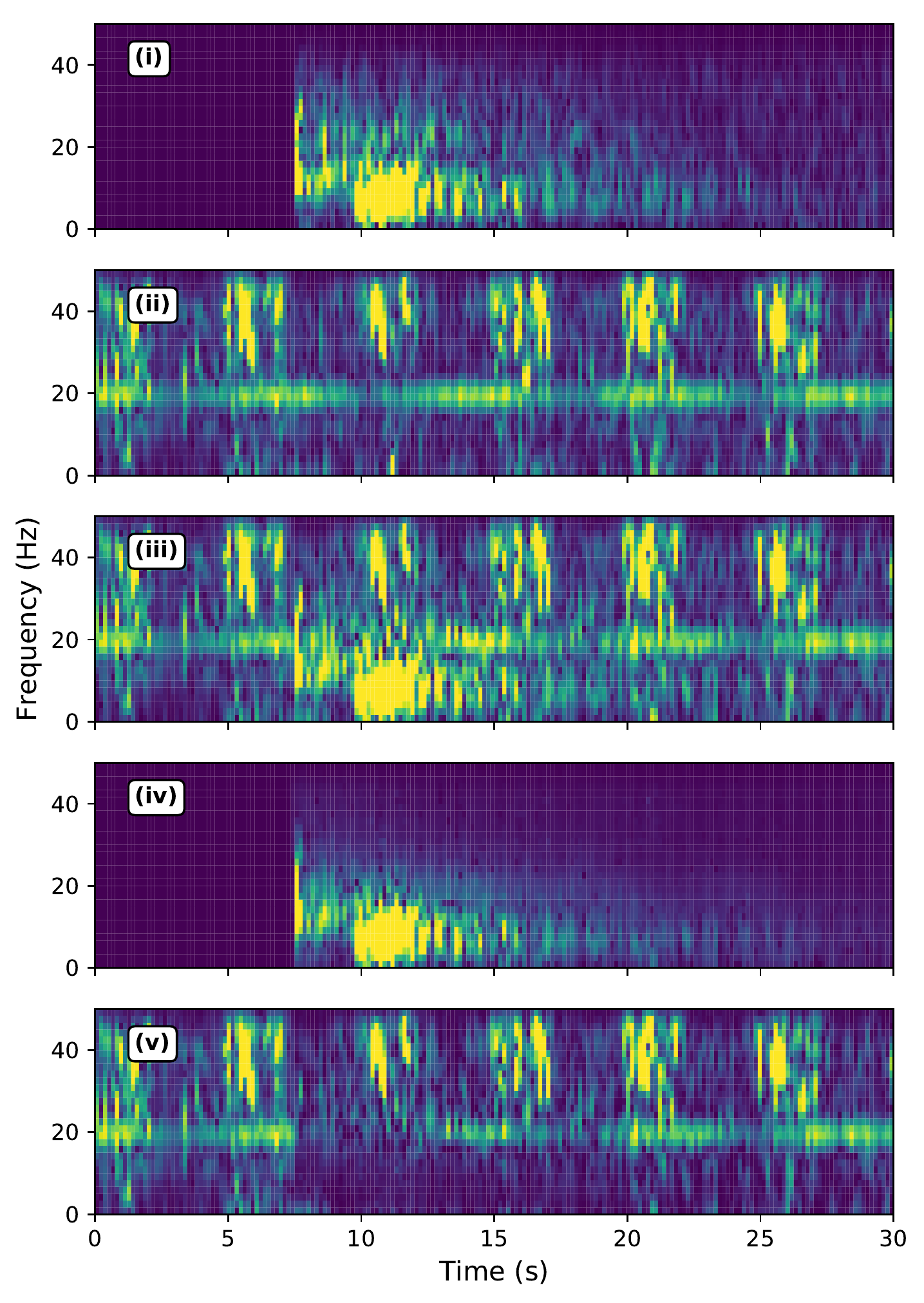}
  \caption{}
\end{subfigure}
\begin{subfigure}{0.41\textwidth}
\centering
  \includegraphics[width=\linewidth]{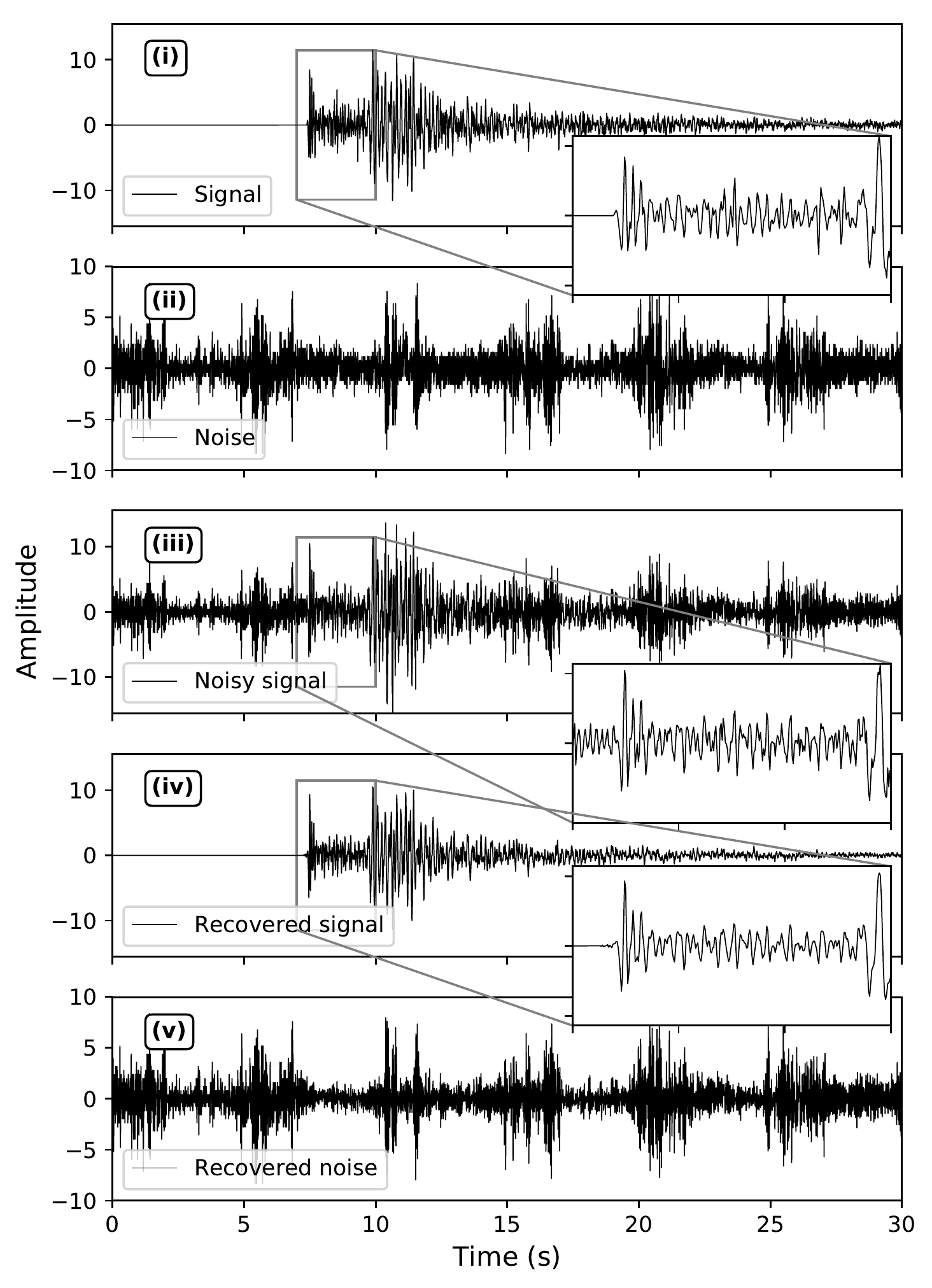}
  \caption{}
\end{subfigure}
\caption{Denoising performance in presence of cyclic non-seismic signals. It is plotted in a same manner as Fig.~\ref{fig:signal01}.}
\label{fig:cycle_noise01}
\end{figure*}

One advantage of our learning-based denoising method is that the network not only learns the features of seismic signals but also the features of various noise. The wide variety of noise sources along with their time-varying signatures makes it very difficult for hand-engineered features to well represent each type of noise. However, DeepDenoiser shows potential ability to learn a sparse feature representation for different types of noise.
Some of them are recognized in our test dataset:
The first type of noise is band-limited (Fig.~\ref{fig:band_noise01}(a, b)), with relatively strong values within narrow frequency bands; 
The second type of noise is low-frequency noise (Fig.~\ref{fig:band_noise01}(c, d))), which presents strong background fluctuations the signal rides on;
The third type of noise is cyclic noise (Fig.~\ref{fig:cycle_noise01}), which is combined with different modes with the frequency band varying with time.
The first two types of noise could be effectively attenuated by band-pass or low-pass filtering given an accurate estimate of the noise frequency bands. The third type of noise is challenging for traditional denoising methods as the noise changes with time and its frequency band overlaps with the frequency band of the target signal.
For all of these types of noise, DeepDenoiser automatically predicts a mask that adapts to the noise features. 
The mask not only estimates the noise frequency bands needed for denoising but also reflects the changes of frequency contents over time. 
The denoised signal and the separated noise (Fig.~\ref{fig:band_noise01}(b, d)(iv, v), Fig.~\ref{fig:cycle_noise01}(b, d)(iv, v)) support that DeepDenoiser has robust denoising performance on various kinds of noise.

\begin{figure*}
\centering
\begin{subfigure}{0.41\textwidth}
\centering
  \includegraphics[width=\linewidth]{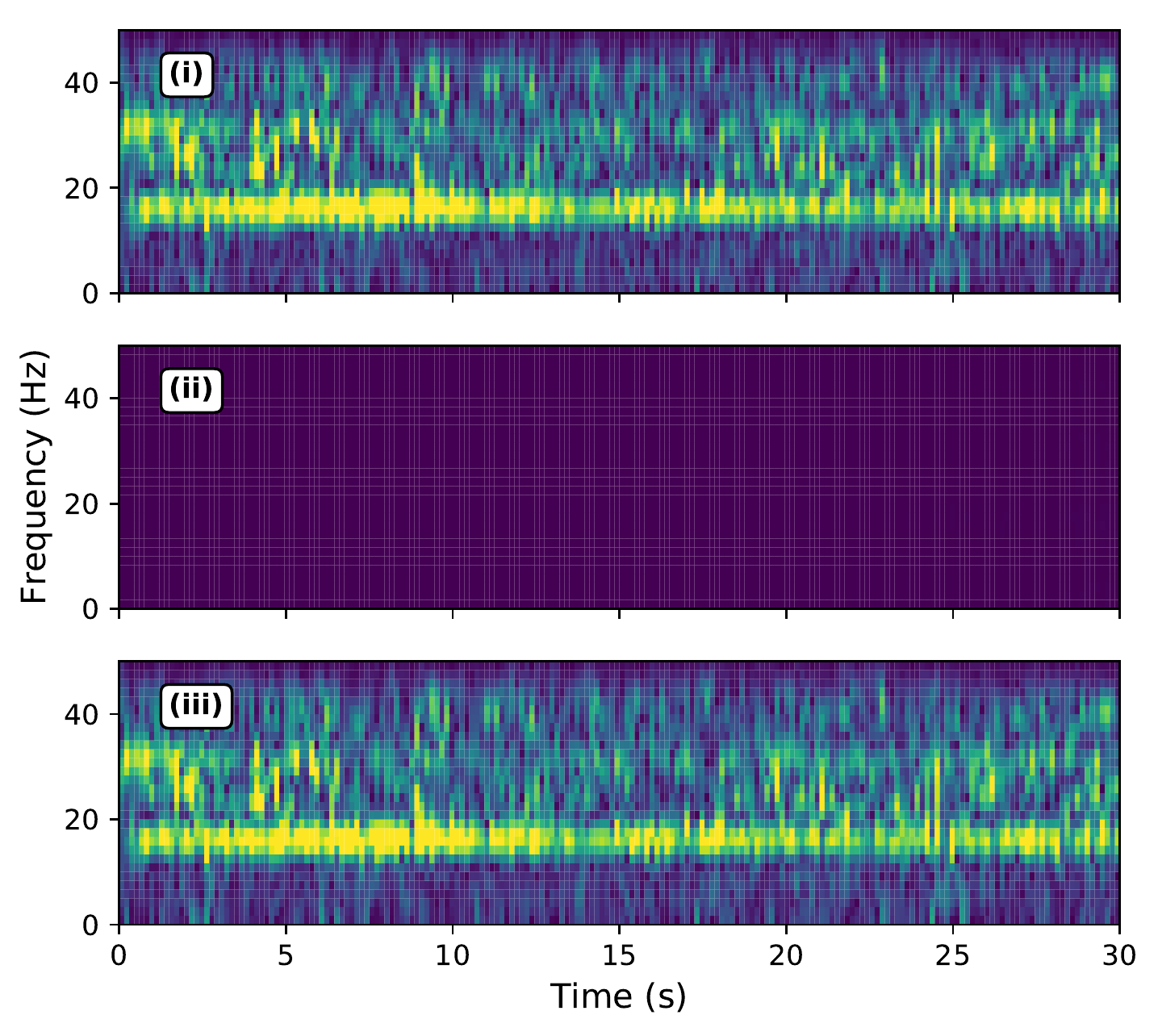}
  \caption{}
\end{subfigure}
\begin{subfigure}{0.41\textwidth}
\centering
  \includegraphics[width=\linewidth]{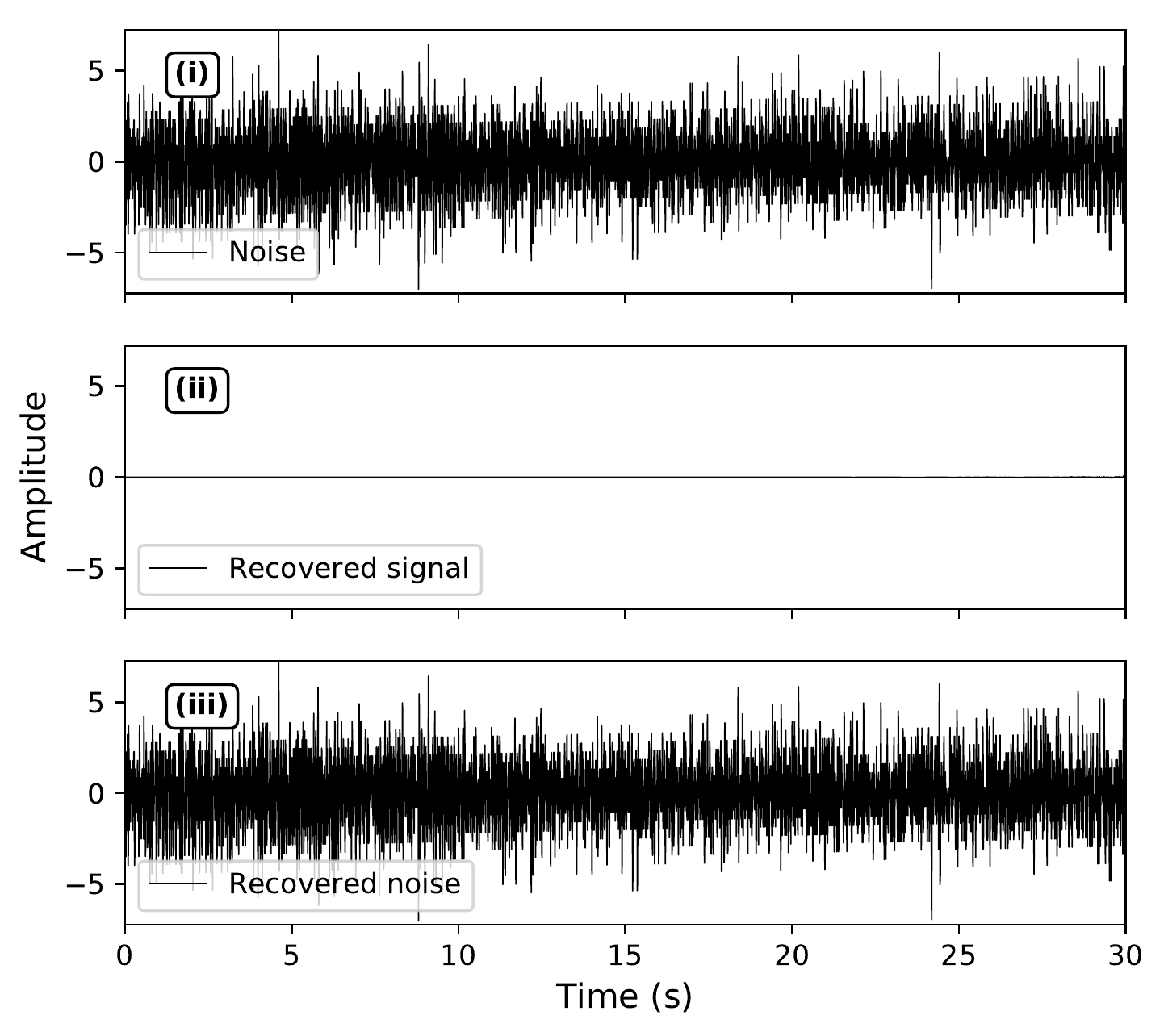}
  \caption{}
\end{subfigure}
\begin{subfigure}{0.41\textwidth}
\centering
  \includegraphics[width=\linewidth]{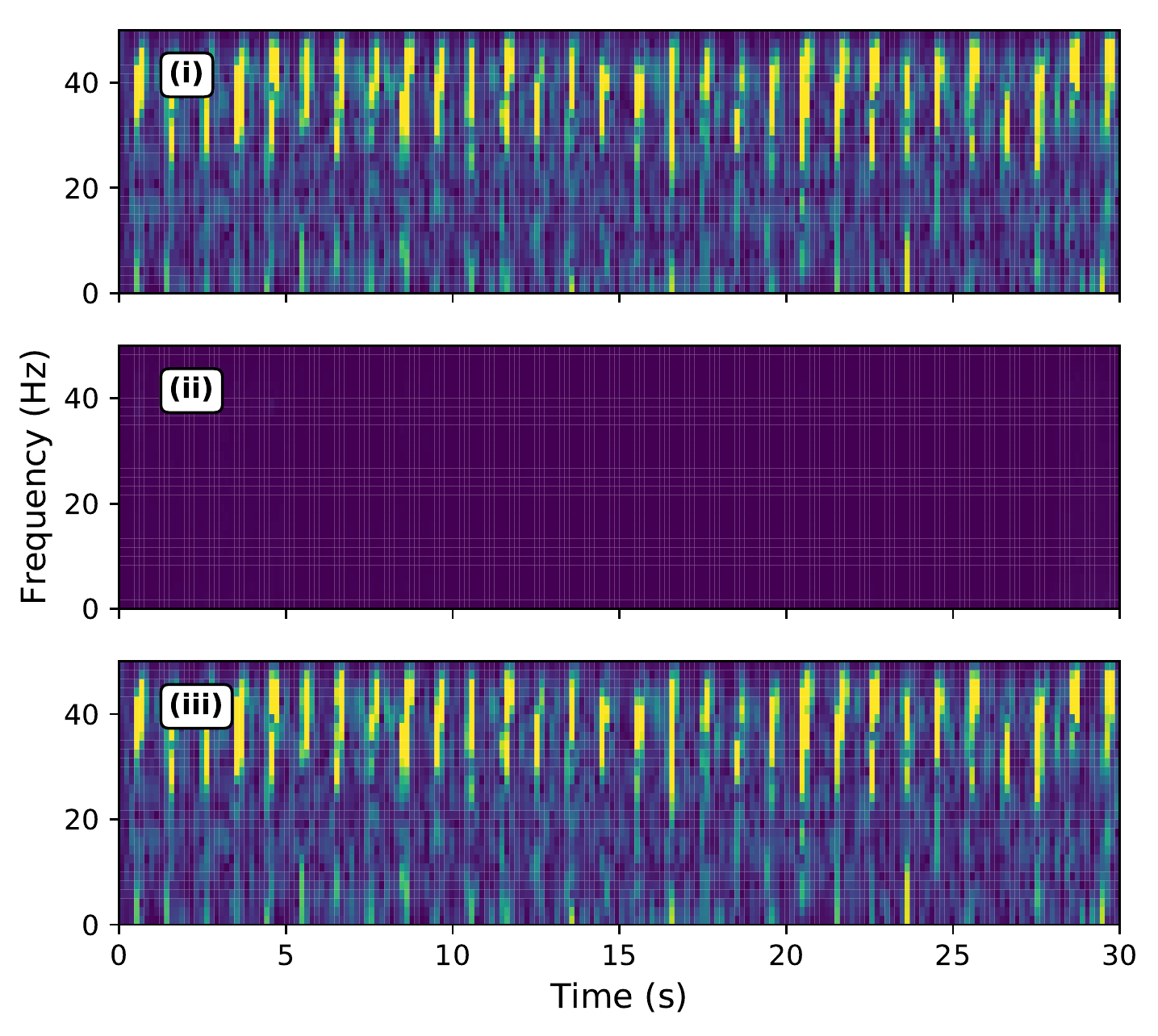}
  \caption{}
\end{subfigure}
\begin{subfigure}{0.41\textwidth}
\centering
  \includegraphics[width=\linewidth]{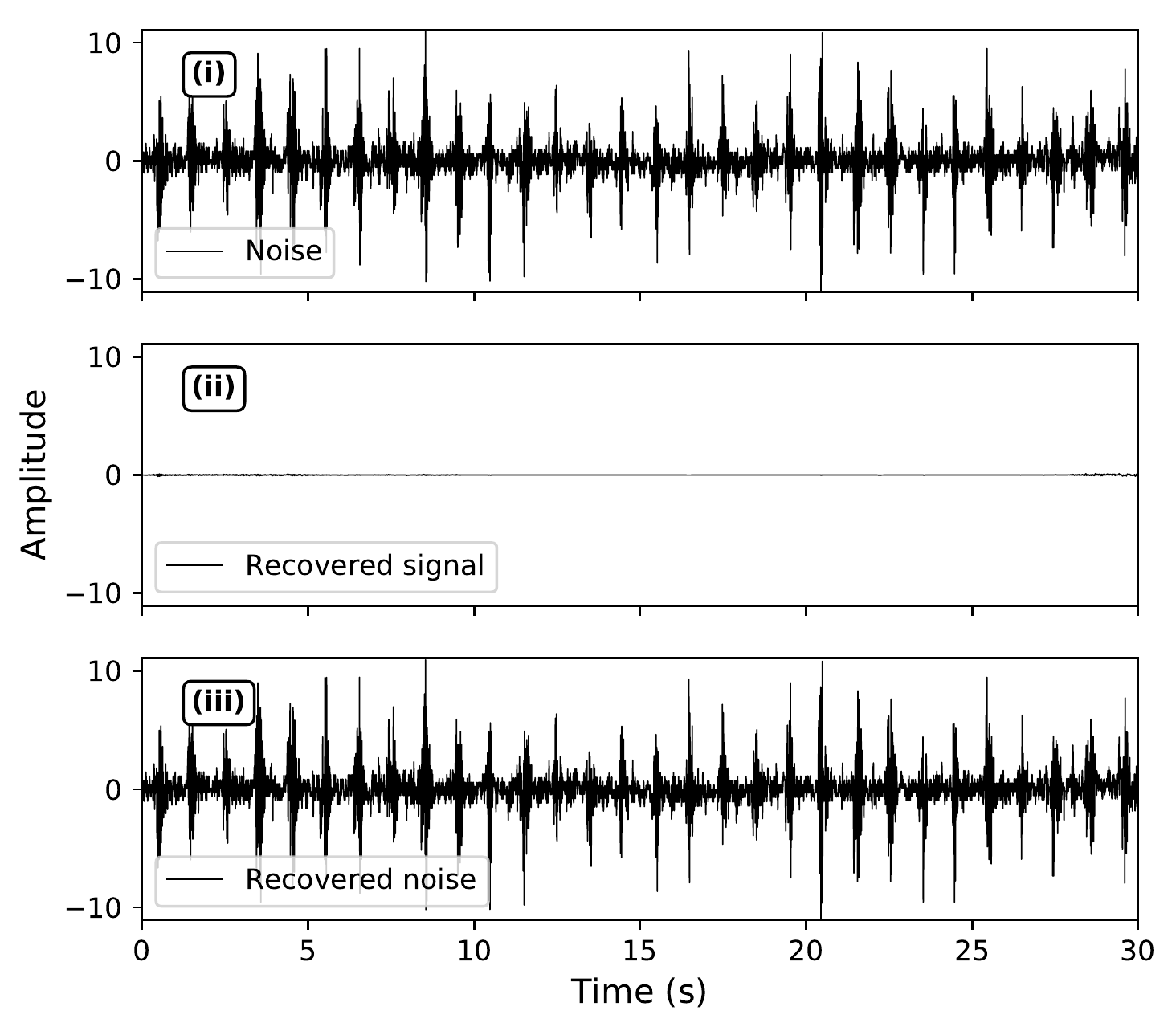}
  \caption{}
\end{subfigure}
\caption{Denoising examples of pure noise: (a, c) time-frequency domain; (b, d) time domain. The inputs are pure noise as shown in panels (i). The predicted signal masks in panels (a, c)(ii) and the recovered signal waveforms in panels (b, d)(ii) are mostly zeros. The predicted noise masks in panels (a, c)(iii) and recovered noise waveforms in panels (b, d)(iii) are same as the input noise in panels (i).}
\label{fig:no_signal01}
\end{figure*}

We also test DeepDenoiser on waveforms with pure noise.
Fig.~\ref{fig:no_signal01} shows that it accurately distinguishes these waveforms from those containing earthquake signals, i.e. no recovered signal is predicted and the recovered noise is equivalent to the input noise. With traditional denoising methods, the input noise waveform on the other hand could be contaminated. This important feature renders DeepDenoiser with the ability to preserve noise signals with or without the presence of earthquakes at reasonable computational cost. It could have significant applications in data preprocessing for ambient seismic noise studies where the contamination of noise data with earthquake signals can heavily affect the cross-correlation results~\citep{liu2016frequency} and the current way of dealing with earthquake contamination is to simply discard the associated windows~\citep{bensen2007processing, sheng2017multicomponent}. 

\subsection{Generalization}
\begin{figure*}
\centering
\begin{subfigure}{0.41\textwidth}
\centering
  \includegraphics[width=\linewidth]{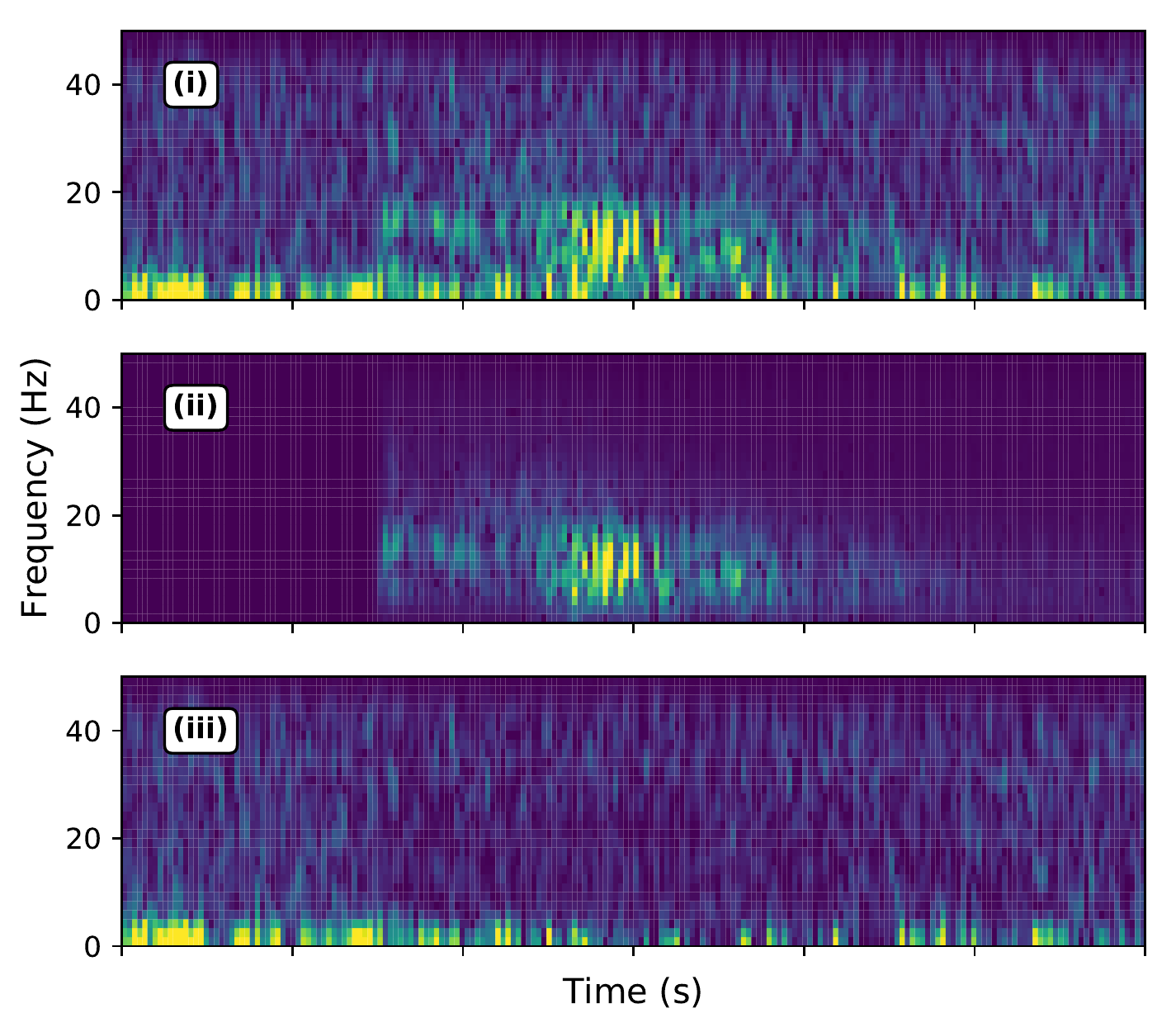}
  \caption{}
\end{subfigure}
\begin{subfigure}{0.41\textwidth}
	\centering
	\includegraphics[width=\linewidth]{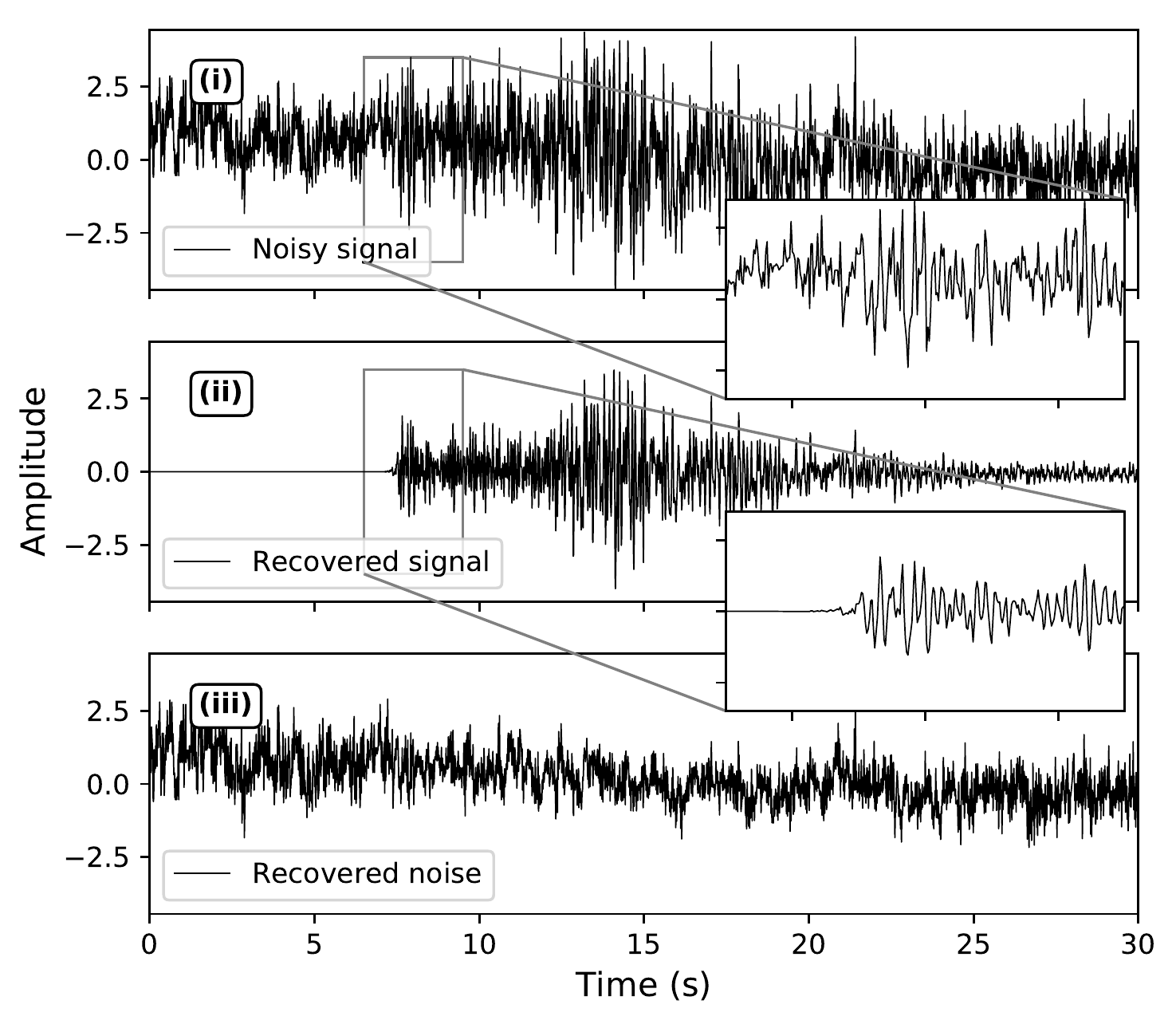}
	\caption{}
\end{subfigure}
\begin{subfigure}{0.41\textwidth}
\centering
  \includegraphics[width=\linewidth]{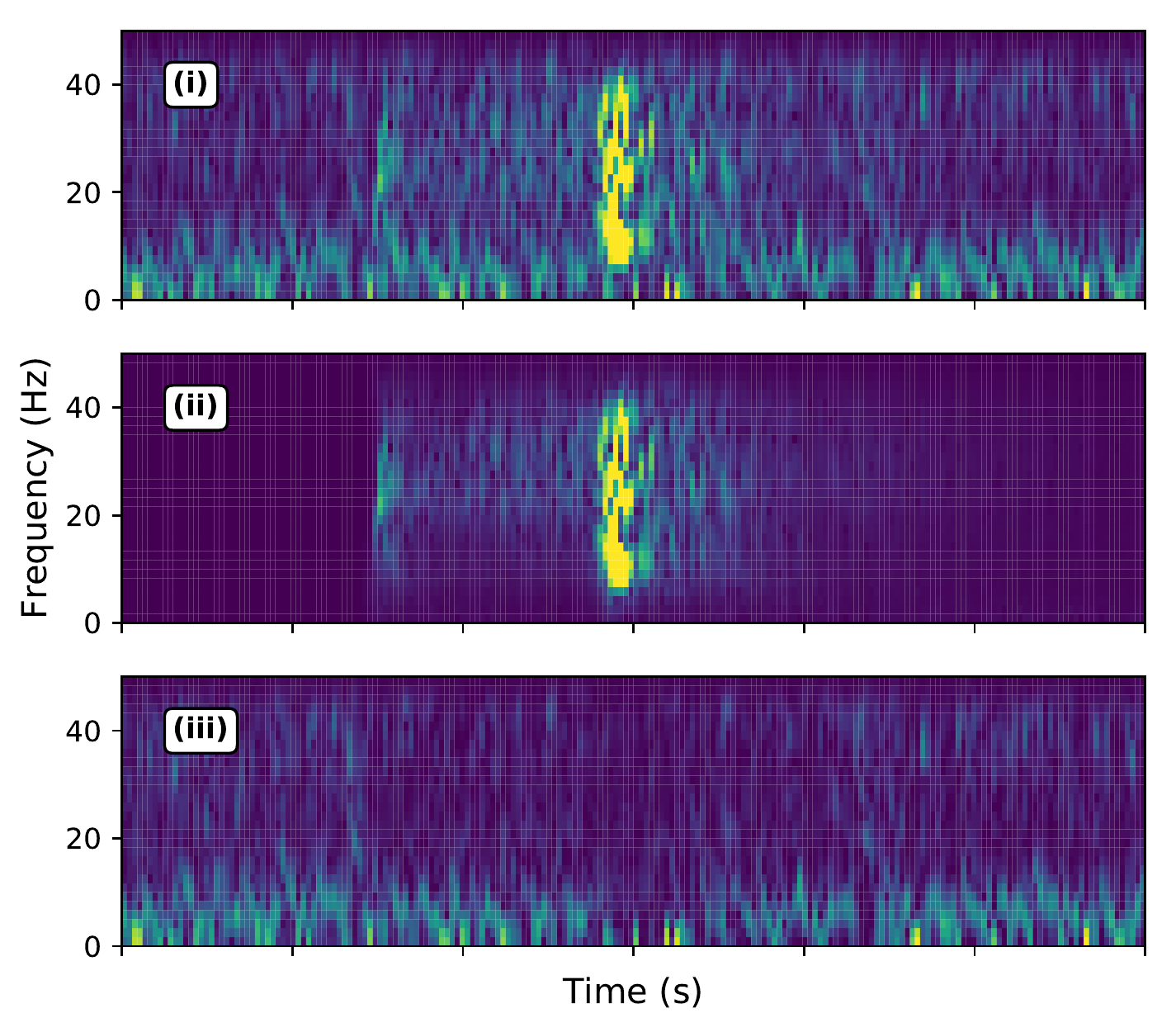}
  \caption{}
\end{subfigure}
\begin{subfigure}{0.41\textwidth}
	\centering
	\includegraphics[width=\linewidth]{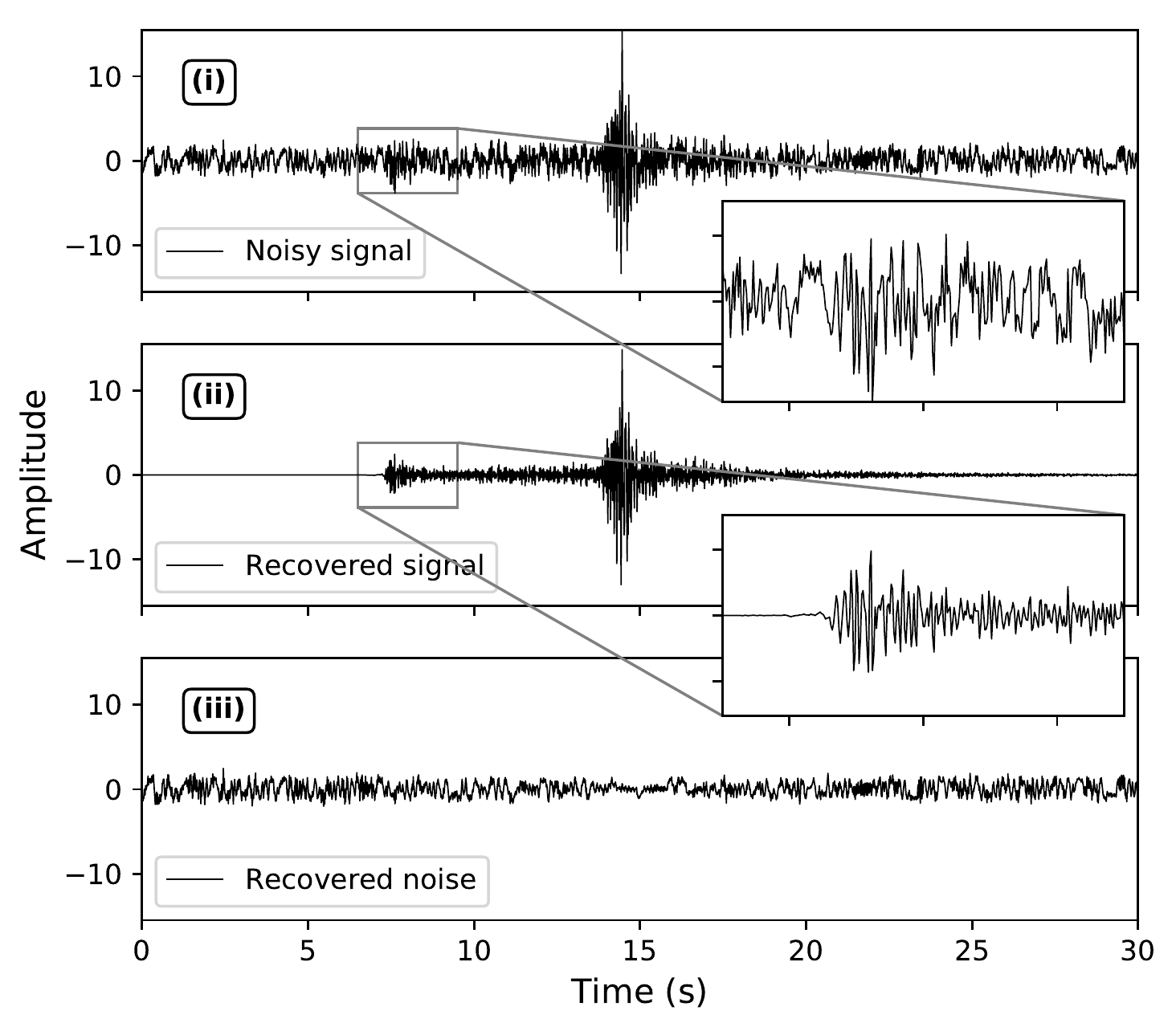}
	\caption{}
\end{subfigure}
\caption{Denoising performance on unseen seismograms: (a, c) time-frequency domain; (b, d) time domain. The time-frequency coefficients and waveforms of the unseen noisy signal are plotted in panels (i). Panels (a, c)(ii) show the denoised signal in the time-frequency domain, and panels (b, d)(ii) show its time domain waveform. The recovered noise is shown in panels (a, c)(iii) and (b, d)(iii).}
\label{fig:real_data01}
\end{figure*}

\begin{figure}
\centering
\includegraphics[width=0.45\textwidth]{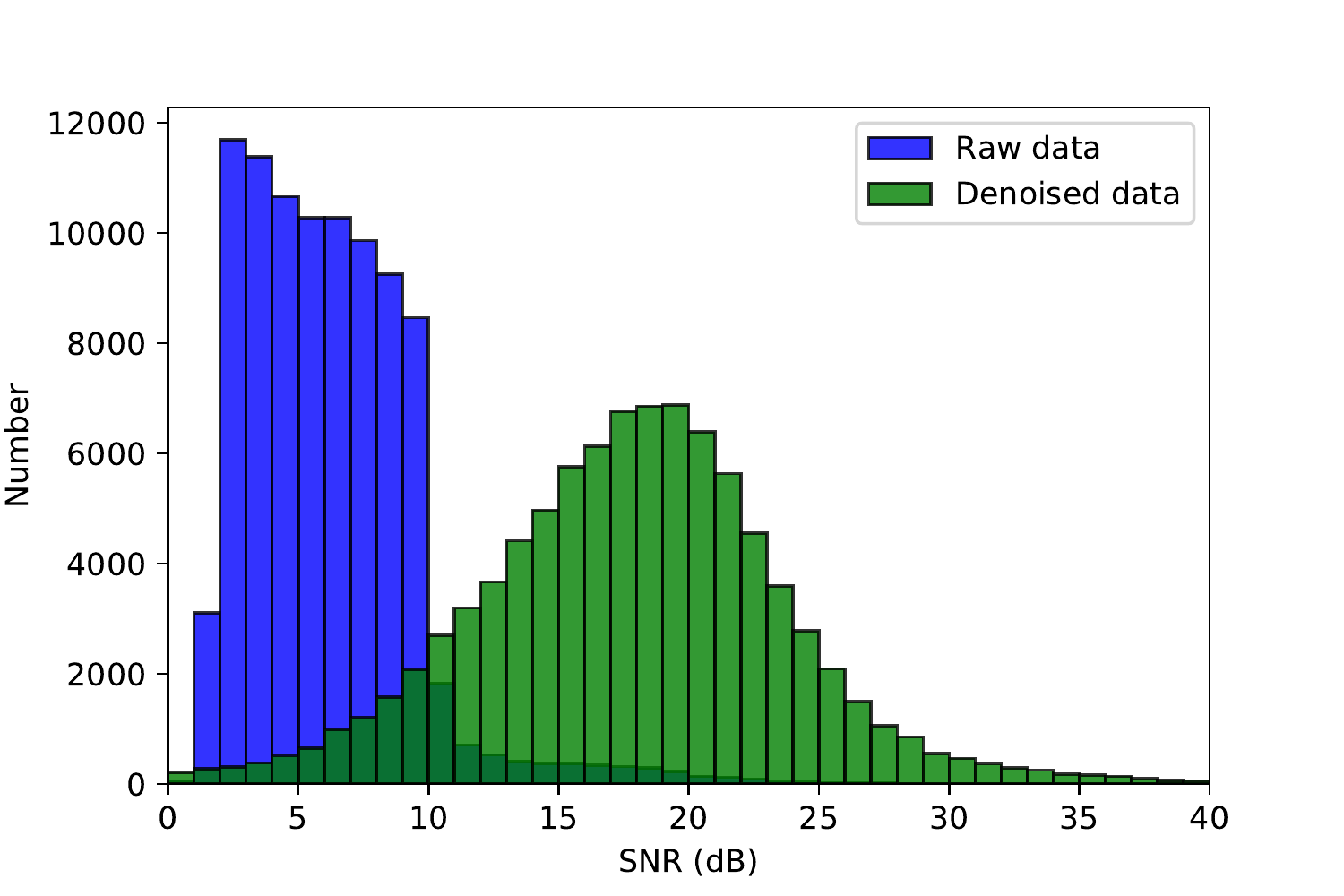}
\caption{Histogram showing SNR improvement of real noisy seismograms at Northern California.}
\label{fig:snr_improvment}
\end{figure}

The neural network is trained on seismic data that is synthesized by superposing real noise signals onto the real high-SNR seismic signals. To test the network's generalizability, we applied it on the 91,000 samples of real seismograms recorded in Northern California. These seismograms are from detected earthquakes in the Northern California Earthquake Catalog, but are contaminated heavily by noise (Fig.~\ref{fig:real_data01}(i)) and therefore have low SNRs (Fig.~\ref{fig:snr_improvment}).

The results suggest that DeepDenoiser successfully recovers clean seismic waveforms (Fig.~\ref{fig:real_data01}(ii)), e.g. the first arrivals, and improves the SNR by around 15 dB (Fig.~\ref{fig:snr_improvment}). The SNR is calculated as:
\begin{equation}
\text{SNR} = 10 \log_{10} \left( \frac{\sigma_\text{signal}}{\sigma_\text{noise}} \right)
\end{equation}
where $\sigma_\text{signal}$ and $\sigma_\text{noise}$ are the standard deviations of waveforms before and after the first arrival.
Although DeepDenoiser is trained on synthetic data, it generalizes well to real seismograms. 
This suggests we can directly apply the deep neural network trained in this study to denoising tasks in real life whose performance relies on clean, undistorted seismic signals.

\subsection{Comparison with Other Methods}
\begin{figure}
\centering
\includegraphics[width=0.45\textwidth]{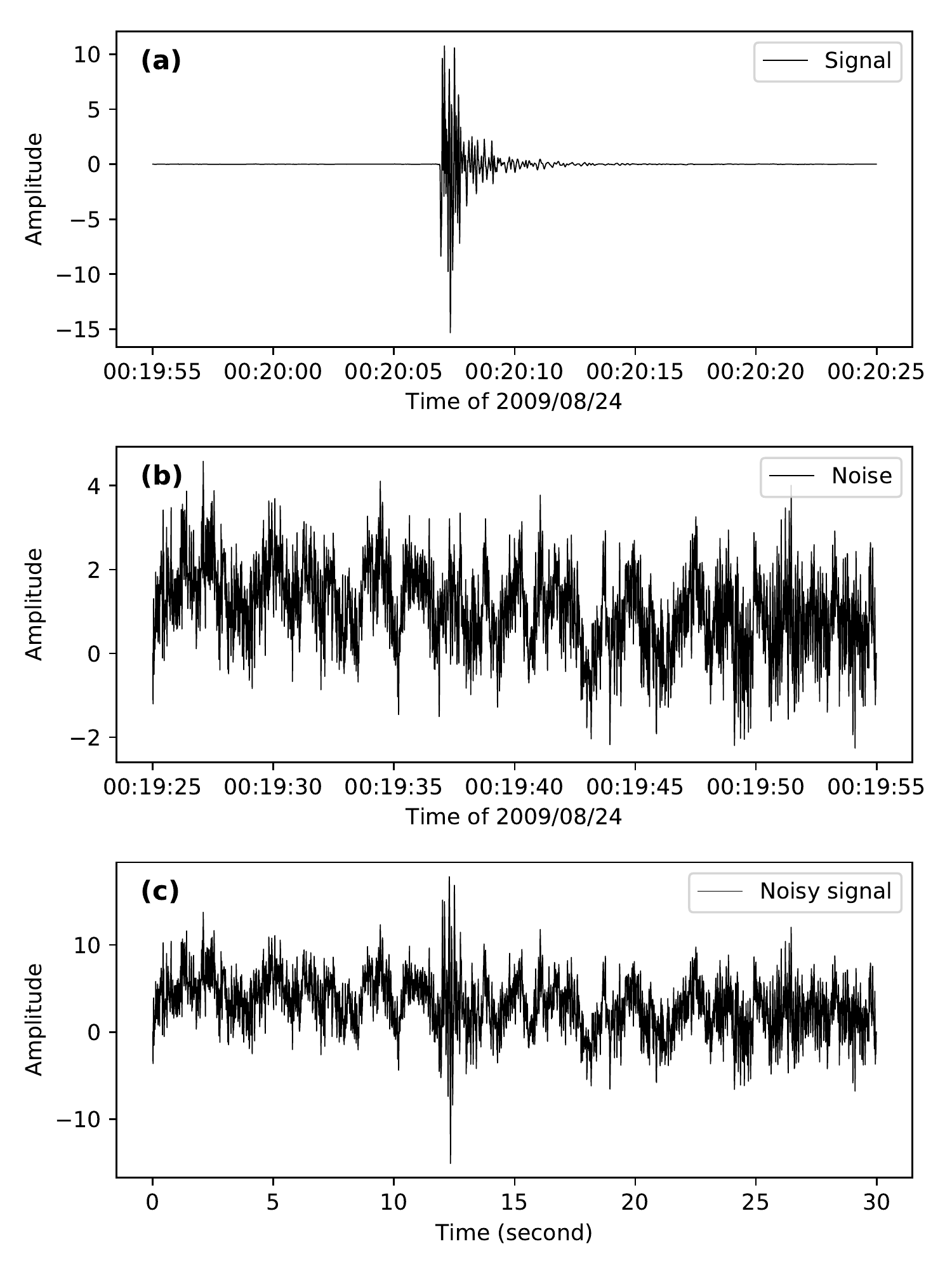}
\caption{The waveforms cut from "EHZ" channel of RTSH station, BayernNetz (BW) Network in Germany: (a) the "clean" signal; (b) the noise waveform obtained from the same station prior to the event origin time; (c) an example of generated noisy waveform with a stacking ratio of 3 for noise in (b) over signal in (a).}
\label{fig:compare_data}
\end{figure}

To compare the performance of our method with other denoising approaches, we select one seismic waveform in "Trigger/Picker Tutorial"\footnote{\url{https://docs.obspy.org/tutorial/code_snippets/trigger_tutorial.html}} of obspy~\citep{Beyreuther2010} as a benchmark. The waveform is from the "EHZ" channel, which represents a short period high gain seismometer, recorded at station RTSH of the BayernNetz (BW) Network in Germany. We also cut a 30-second window prior to the event as the noise sample (Fig.~\ref{fig:compare_data}). We scale the noise waveform to vary the noise level and stack it with the seismic waveform to generate a sequence of noisy signals with varying SNRs (Fig.~\ref{fig:compare_data}(c)).

\begin{figure}
\centering
\includegraphics[width=0.45\textwidth]{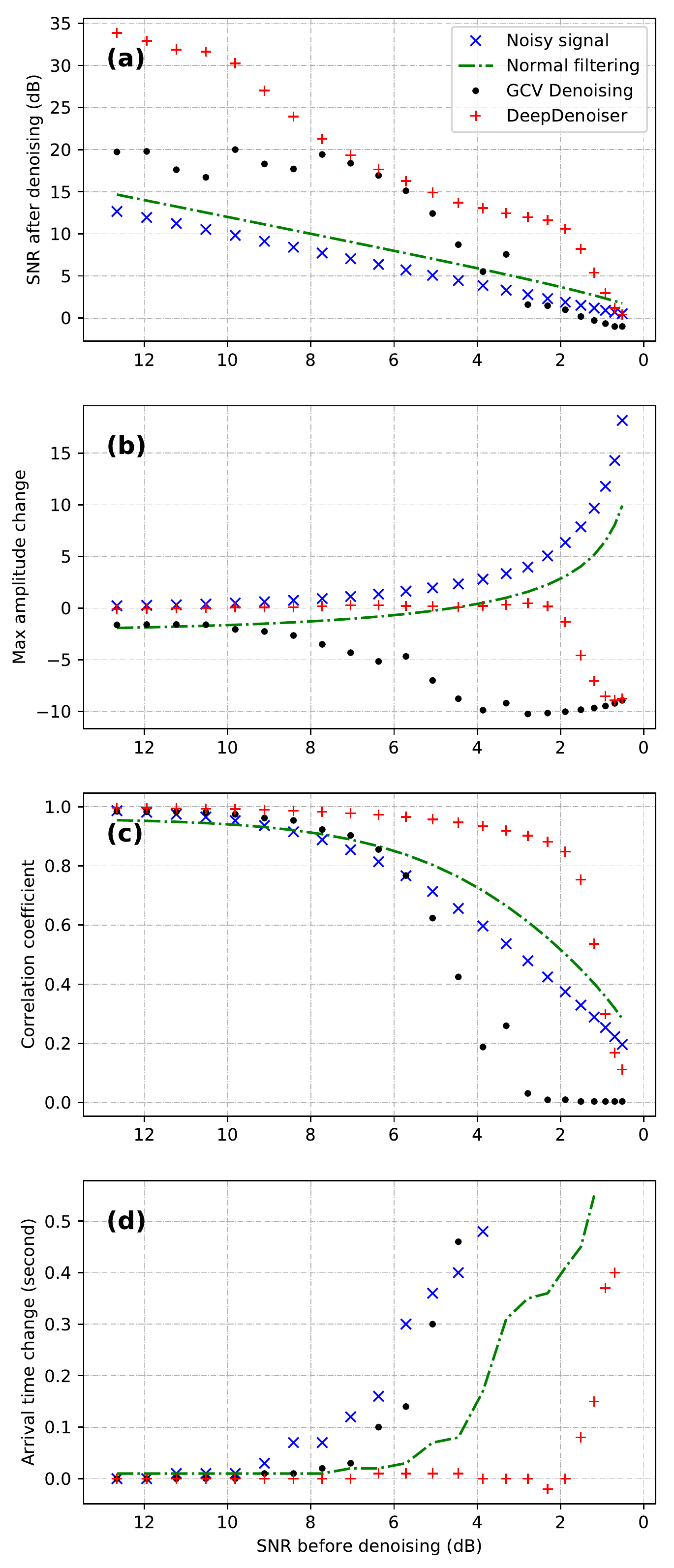}
\caption{Performance comparison between normal filtering, GCV denoising and DeepDenoiser: (a) improvement of SNR; (b) max amplitude changes; (c) correlation coefficient; (d) picked arrival time changes. Values in (b) and (d) are differences compared with the signal in Fig.~\ref{fig:compare_data}(a). Values in (c) are calculated by zero-lag cross-correlation between the denoised waveforms and the signal in Fig.~\ref{fig:compare_data}(a).}
\label{fig:compare_result}
\end{figure}

We used normal filtering and general cross-validation denoising~\citep{Mousavi2017CGV} for the comparison. For normal filtering, we design the filter based on the smoothed frequency distribution of the clean signal (Fig.~\ref{fig:compare_filter} in appendix), which is intended to emphasize the frequency band where the true signal resides. We measure the SNR improvements between the denoised and noisy signals, and changes of the maximum amplitude,  the correlation coefficient and the picked arrival times as the bases of the comparison between the performance of these methods (Fig.~\ref{fig:compare_result}).
The arrival time is picked using the same STA/LTA method in the section below.

Fig.~\ref{fig:compare_result} indicates that DeepDenoiser achieves a better denoising performance while introducing smaller distortion to the signal waveform.
The SNR improvement of DeepDenoiser is more significant and more robust than the GCV method (Fig.~\ref{fig:compare_result}(a)).
Fig.~\ref{fig:compare_result}(b) shows that DeepDenoiser recovers the true amplitude of the signal more accurately.
The max amplitude of the denoised signal is closer to the true signal in Fig.~\ref{fig:compare_data} when the SNR is larger than 2 dB, while the GCV method largely attenuates both noise and signal to achieve a good denoising performance.
The higher correlation coefficients in Fig.~\ref{fig:compare_result}(c) further demonstrates the DeepDenoiser introduces smaller waveform distortion during denoising. In contrast, the GCV method distorts the signal waveform significantly.
DeepDenoiser is designed based on a way to separate signal and noise effectively, which enables it to improve SNR and preserve the signal waveform simultaneously.
The denoised waveform also improves the recovery of arrival times from noisy signal. With a fixed activation threshold for STA/LTA method, the picked arrival times will have an sytematically increasing  delay and eventually not be recovered at increasing noise levels. Fig.~\ref{fig:compare_result}(d) shows that on the noisy signal the arrival time has a delay of 0.5 s at SNR of 4 dB and fails to be recovered beyond this level (smaller dB). On the other hand, the denoised signal by DeepDenoiser yields an accurate arrival time even at SNR of 2dB, which is also better than the other two denoising methods.

\subsection{Application for Earthquake Detection}

\begin{figure*}
\begin{subfigure}{\textwidth}
\centering
\includegraphics[width=0.9\linewidth]{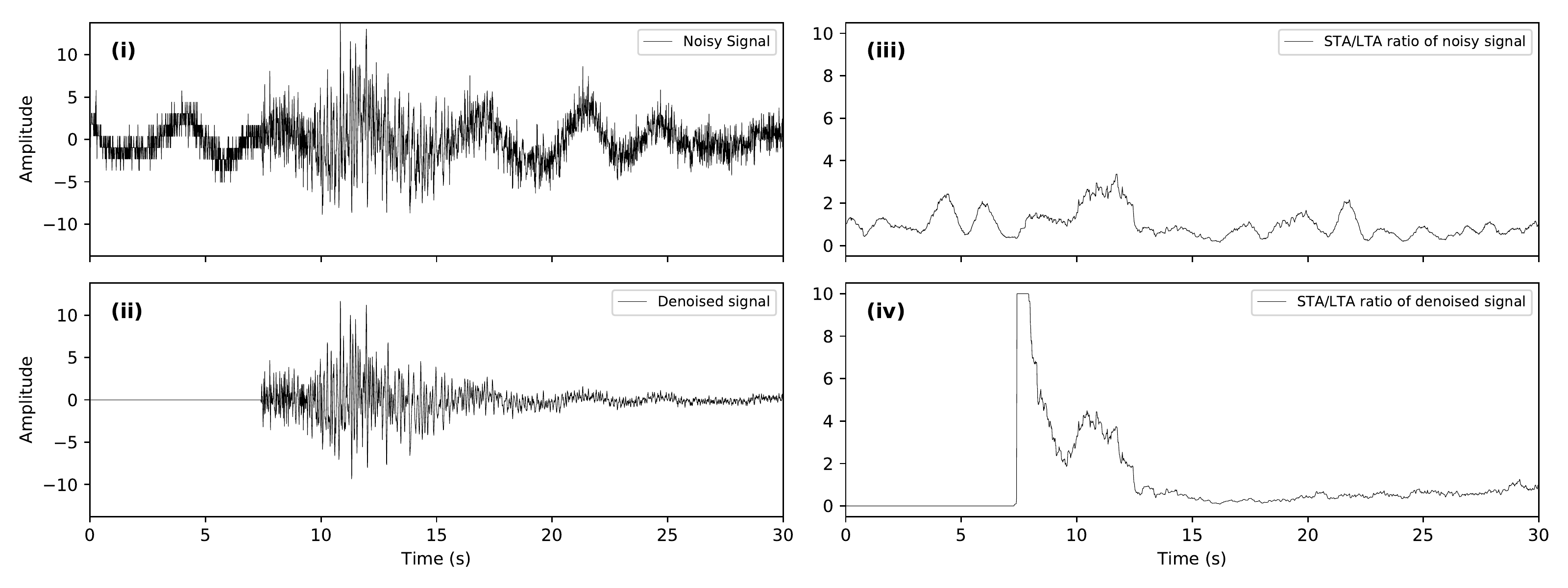}
\caption{}
\end{subfigure}
\begin{subfigure}{\textwidth}
\centering
\includegraphics[width=0.9\linewidth]{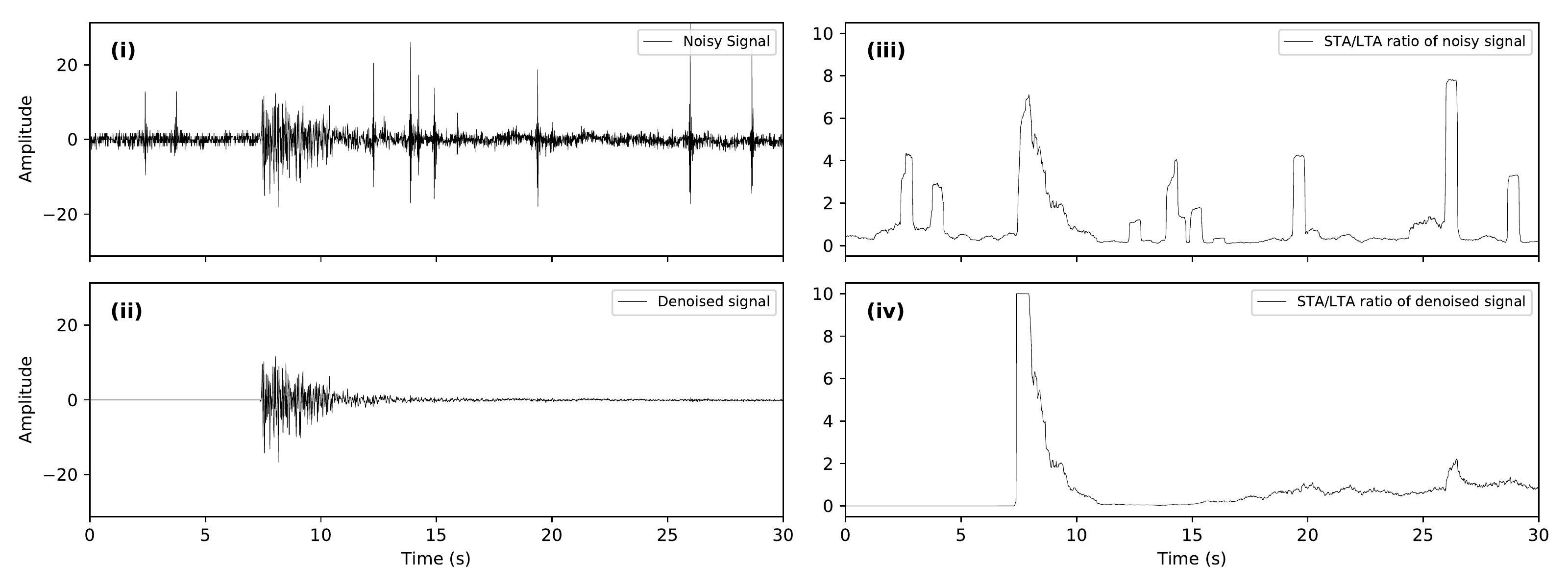}
\caption{}
\end{subfigure}
\caption{Improvement of STA/LTA characteristic function after denoising: (i) noisy signals; (ii) denoised signals; (iii) and (iv) the corresponding STAT/LTA characteristic functions.}
\label{fig:stalta01}
\end{figure*}

Background noise can significantly affect the performance of common detection algorithms such as STA/LTA (short-term averaging/long-term averaging) for detecting small and weak events~\citep{withers1998comparison}. Moreover, the presence of non-earthquake signals will increase the false positive rate, and as a result, degrade detection precision. 
In Fig.~\ref{fig:stalta01} we present two examples of how DeepDenoiser can improve the STA/LTA characteristic function. The short and long time windows are set to be 0.5 seconds and 5 seconds.
Fig.~\ref{fig:stalta01}(a) shows an example when the noise smooths out the sharp jump at the arrival of seismic waves and makes the earthquake undetectable. DeepDenoiser makes this earthquake easy to detect and increases the recall rate by making the STA/LTA characteristic function sharp after denoising. 
Fig.~\ref{fig:stalta01}(b) shows another example when impulsive non-seismic signals bring sharp peaks to the  STA/LTA characteristic function. These peaks will be falsely detected as earthquakes by the STA/LTA method. DeepDenoiser can effectively remove these non-seismic signals and increase the prediction precision. 
We compare the earthquake detection results on our test dataset of 10,800 samples before and after denoising. With a threshold of 5, we calculate the precision of earthquake detection to be 35.14$\%$, 92.34$\%$ and the recall rates to be 17.69$\%$, 93.76$\%$ for the noisy signal and denoised signal respectively. 
The accuracy is defined as the ratio of true positive detections over the total positive detections. The recall rate is defined as the ratio of true positives over the total true number of earthquakes.
Both the accuracy and recall rate are improved significantly by DeepDenoiser.

\section{Discussion and Conclusion}

We have developed a novel deep-learning-based denoising algorithm for seismic data, DeepDenoiser. This learning-based approach can learn a collection of sparse features with the aim of signal and noise separation from samples of data. These features reflect more accurately the statistical characteristics of a signal of interest and can be used for effective denoising or decomposition of input waveforms. The neural network automatically determines the percentage of signal present at each data point in the time-frequency space (mask). Learning a sparse representation of data and predicting two adaptive thresholding masks is performed simultaneously by optimizing the loss function. The masks determined by the network are then used to effectively decompose the input data into a signal of interest and noise. 

Our results show this algorithm can achieve robust and effective performance in denoising of data even when the signal and noise share a common frequency band. The denoising ability of our method is not limited to random white noise, but performs well for a variety of colored noise and non-earthquake signals as well. Our tests indicate that DeepDenoiser can significantly improve the SNR with minimal change to the underlying signal of interest. DeepDenoiser preserves waveform shape more faithfully than other denoising methods, even in presence of high noise levels. The network appears to be able to generalize to datasets outside of training set. We have only demonstrated the capability of our method in improving event detection; however, the potential applications of our approach are widespread. DeepDenoiser can be adapted for various types of seismic and non-seismic signals and different applications. Seismic imaging, micro-seismic monitoring, test-ban treaty monitoring, and preprocessing of ambient noise data are other potential applications of our method.

Although the performance of DeepDenoiser is impressive, it does not result in a perfect separation of signal and noise. Perfect separation of signal and noise in the time-frequency domain requires recovering two complex vectors for signal and noise from the complex vector of noisy signal, while DeepDenoiser uses the same mask for both the real and imaginary parts, and the mask, which has a value between [0, 1], can not recover  signal values larger than the input noisy signal. Predicting the signal and noise values directly without the use of a mask may be a future direction for improvement.

\section*{Acknowledgments} 
We thank Lind S. Gee and Stephane Zuzlewski for their help on downloading and processing the catalog and waveform data from NCEDC. We thank William Ellsworth, Kaiwen Wang, Yixiao Sheng, Kai Sheng Tai and Kexin Rong for helpful discussions. Waveform data, metadata, or data products for this study were accessed through the Northern California Earthquake Data Center (NCEDC). This research is supported by the National Science Foundation (NSF) grant number EAR-1818579.

\appendices
\section{}
\begin{figure}
\centering
\includegraphics[width=0.45\textwidth]{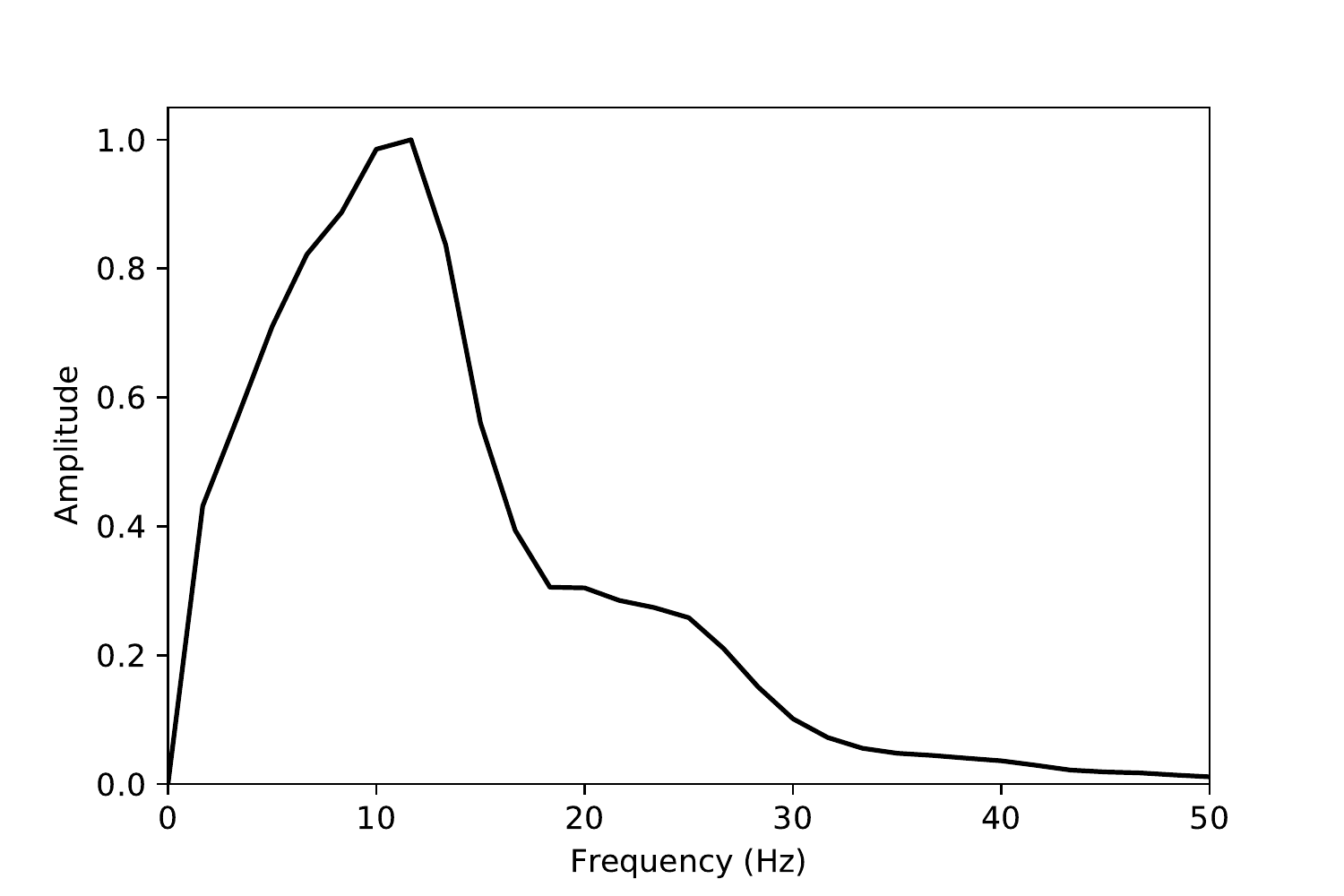}
\caption{Frequency distribution of the filter used for normal filtering in the comparison section}
\label{fig:compare_filter}
\end{figure}

Frequency distribution of the filter used for normal filtering in the comparison section are shown in Fig.~\ref{fig:compare_filter}. This is built based on the frequency distribution of clean signal in Fig.~\ref{fig:compare_data}(a), so that it could keep the frequency band of the clean signal. 

\bibliographystyle{IEEEtranN}
\bibliography{denoising.bib}

\end{document}